\documentclass{aa}  
\usepackage{graphicx}
\usepackage{natbib}
\bibpunct{(}{)}{;}{a}{}{,} 
\usepackage{lscape}
\usepackage{longtable}
\usepackage{lscape}
\usepackage{color}
\usepackage{rotating}
\usepackage[percent]{overpic}
\usepackage{multirow}
\usepackage[per-mode=symbol, 
			separate-uncertainty, 
			multi-part-units=single, 
			list-units=single, 
			range-units=single, 
						load-configurations=abbreviations,
						]
						{siunitx}
\DeclareSIUnit\arcsec{arcsec}
\DeclareSIUnit\mag{mag}
\DeclareSIUnit\gal{gal}
\usepackage{txfonts}

\begin{document}
\title{Statistical analysis of bound companions in the Coma cluster}

   \author{Martin Mendelin\inst{}
          \and
          Bruno Binggeli\inst{}
          }

	\institute{Department of Physics, Basel University, Klingelbergstrasse 82, 4056 Basel, Switzerland\\				
	\email{martin.mendelin@intergga.ch; bruno.binggeli@unibas.ch}}

   \date{Received 6 February 2017; accepted 12 April 2017}

 
  \abstract {} {The rich and nearby Coma cluster of galaxies is known to have substructure. We aim to create a more detailed picture of this substructure by searching directly for bound companions around individual giant members.} 
{We have used two catalogs of Coma galaxies, one covering the cluster core for a detailed morphological analysis, another covering the outskirts. The separation limit between possible companions (secondaries) and giants (primaries) is chosen as $M_B = -19$ and $M_R = -20$, respectively for the two catalogs. We have created pseudo-clusters by shuffling positions or velocities of the primaries and search for significant over-densities of possible companions around giants by comparison with the data. This method was developed and applied first to the Virgo cluster. 
In a second approach we introduced a modified nearest neighbor analysis using several interaction parameters for all galaxies.} 
{We find evidence for some excesses due to possible companions for both catalogs. Satellites are typically found among the faintest dwarfs ($M_B < -16$) around high-luminosity primaries. The most significant excesses are found around very luminous late-type giants (spirals) in the outskirts, which is expected in an infall scenario of cluster evolution. A rough estimate for an upper limit of bound galaxies within Coma is $\sim 2 - 4\%$, to be compared with $\sim 7\%$ for Virgo.} 
{The results agree well with the expected low frequency of bound companions in a regular cluster such as Coma. To exploit the data more fully and reach more detailed insights into the physics of cluster evolution we suggest applying the method also to model clusters created by $N$-body simulations for comparison.}

   \keywords{Galaxies: clusters: individual: Coma - galaxies: dwarf - galaxies: interactions - galaxies: statistics - galaxies: structure
               }

   \maketitle
   
%

\section{Introduction}
\label{sec:Intro}
In the $\Lambda$CDM standard scenario of structure formation, clusters of galaxies form hierarchically by the accretion and subsequent merging of smaller units in a bottom-up manner \citep[e.g., the reviews of][]{Planelles,KraBor}. As the merging and virialization of subcluster units is a slow process and new groups of galaxies are arriving and falling into a cluster from the environment even at the present cosmic epoch \citep[][and references therein]{TulShay, A05, Co12}, this process has left its traces by the simple presence of substructure in most clusters.
The degree of substructure present, along with the mixture of galaxy types, is therefore taken as a measure of the evolutionary stage of a cluster. The less substructure present the more evolved and relaxed a cluster of galaxies is. A very rough morphological classification proposed by \citet{Ab75}, following \citet{Zw61} and \citet{Mo61}, distinguishes between `regular', mostly rich clusters showing little substructure but a high degree of concentration and symmetry with few spiral galaxies, and `irregular', mostly poor clusters with a high degree of substructure and asymmetry containing many spiral galaxies (see also the classification schemes of \citet{BauMo71}, and \citet{RoSa67} which capture more morphological details but are based on the same basic distinction). The prototypes of a regular and a irregular cluster are the nearby, best-studied Coma cluster and Virgo cluster, respectively.  

However, even the rich, regular \& relaxed Coma cluster (A1656), which is in the focus of the present paper, shows significant substructure (\citeauthor{Biviano96}, \citeyear{Biviano96}; \citeauthor{BivianoRev}, \citeyear{BivianoRev};
\citeauthor{CD96}, \citeyear{CD96}, hereafter CD96;
\citeauthor{N03}, \citeyear{N03};
\citeauthor{A05}, \citeyear{A05}, hereafter A05). The basic substructure of Coma is its binarity induced by the two dominant giant galaxies NGC 4889 and 4874. However, as much as 17 subgroups were identified in Coma by A05 using the hierarchical method of \cite{SernaGerbal}. 

There is a host of sophisticated statistical methods for the analysis of substructure in clusters of galaxies \citep[e.g.,][and references in both]{WB13,Y15}. An alternative, rather simple approach to substructure, though applicable only to relatively nearby, well cataloged clusters, is to look directly for bound companions around massive cluster galaxies. 
In the plausible infall scenario of cluster formation (see above), the infalling groups would consist of a small number of giant galaxies that are surrounded by swarms of bound dwarf galaxy satellites — a phenomenon best known from the Local Group. After the infall, most of these satellites would get stripped off (`liberated') and henceforth move freely in the cluster potential. But a small fraction of the dwarfs, presumably depending (among other things) on the mass and type of the mother galaxy, is expected to survive as companions. 

An elegant statistical method to look for bound companions in clusters was developed and applied to the Virgo cluster by \citet[hereafter F92]{F92}. It is the purpose of the present paper to apply \citeauthor{F92}'s (\citeyear{F92}) method to the Coma cluster, the second-best cataloged rich cluster which is clearly of a different type than Virgo. In a nutshell \citeauthor{F92}'s method works like this: First the cluster sample is divided into a small sample of `primaries' (galaxies brighter than $M$ = -19) and a larger sample of `secondaries' (galaxies fainter than $M$ = -19). Then the number of secondaries is counted within a step-wise growing distance from each primary. These numbers have to be compared with the statistically expected numbers of apparent companions produced by projection effects. This is achieved by repeating the counting process for a large number of Monte Carlo clusters where the positions of the primaries are randomly changed along circles around the cluster center. Any excess of the observed numbers of secondaries around primaries over the corresponding mean numbers from the pseudo-clusters would then indicate the presence of gravitationally bound companions around the primaries. This procedure can be performed as a function of (primary and secondary) luminosity, morphological type and velocity. There are further refinements possible, for example by the introduction of a `interaction parameter' (see F92, also below). \citeauthor{F92}'s procedure bears some resemblance to the two-point cross correlation function but has the advantage that the information about the primaries’ location in the cluster is not lost. Compared to the \cite{SernaGerbal} method it is also possible to give a confidence level for the presence of bound satellites. The density excesses due to bound companions found by \citet{F92} are small but significant, \citeauthor{F92} estimated a minimum of 7\% of cluster members to be bound companions in the Virgo cluster. 

\citeauthor{F92}'s (\citeyear{F92}) work found surprisingly little follow-up. \citet{CWG95} used his procedure to look for pairs in \citeauthor{D80A}'s (\citeyear{D80A}) clusters, but there is so far no \citeauthor{F92}-type analysis of another morphologically resolved nearby cluster. Other very nearby clusters like the Ursa Major and Fornax clusters at roughly Virgo distance lack the richness warranted for a \citeauthor{F92}-type study. So the next target for such a study is clearly Coma. There is in fact some previous work on dwarf companions in the Coma cluster due to \citet{SHP97}. These authors studied the distribution of faint dwarfs around a number of early-type giants in a restricted area covering the cluster core and found on average 4$\pm$1 objects per giant in excess of the local background. However, there is meanwhile good coverage of the whole cluster to a decent depth to allow a full-fledged \citeauthor{F92} analysis of Coma to be persued.       

The results for Coma are expected — and indeed are shown here — to differ significantly from those for Virgo due to the different population and dynamical state of the two clusters. The unrelaxed Virgo cluster exhibits a higher fraction of late-type members than the more virialized, denser Coma cluster \citep[e.g., the review of][]{Boselli}, which is part of the well-known morphology-density relation of galaxies \citep{D80A,D80B}. Infalling groups seem to be preferentially centered on luminous spirals. Thus, one may expect to find bound satellites in clusters preferentially around spirals or in subgroups and not around individual early-type giants. Spirals generally exhibit a higher number of companions than E’s and S0’s, as shown for example by \cite{BothunSull}. Even more important concerning a systematic difference between the clusters, we expect that individual primaries, in particular the old members of of early type, have lost their satellites in the dense environment of Coma due to interactions with the cluster potential as a whole or with other individual galaxies (by tidal interaction, harassment etc., see \citeauthor{Boselli}, \citeyear{Boselli}). In principle, the bound companions, instead of being survivors from infall, not having been ripped off their mother galaxy, could also have been captured later on in the cluster, especially by very massive cluster galaxies. However, numerical simulations indicate that this capture process is at best secondary \citep{BMP99}. 

There are many processes in a cluster environment that tend to transform galaxies from late-type to early-type morphology (giants and companions alike) and systematically wipe out subclustering, including the presence of bound dwarf companions. Our simple statistical analysis, while providing a lot of morphological details with the observed subclustering, is not able to distinguish these processes. 
The purpose of the present paper is primarily to carry out a \citeauthor{F92}-type analysis for the Coma cluster in as much detail as possible, to have a comparison with \citeauthor{F92}'s (\citeyear{F92}) original work on the Virgo cluster.

The rest of the paper is organized as follows: In Sect.~\ref{sec:samples} the catalogs used are presented. Sect.~\ref{sec:Method} deals with the applied MC methods, the computation of the projected secondary density around primaries $\Sigma(r)$ and the interaction parameter. The results are presented in Sect.~\ref{sec:RESULTS} and discussed in Sect.~\ref{sec:discussion}. A short conclusion and prospect are given in Sect.~\ref{sec:concl}.


\section{Samples}
\label{sec:samples}
While the Coma cluster has been studied extensively for the last few decades, there does not yet exist such a complete catalog as the VCC by \cite{VCC} for the Virgo cluster. Nonetheless, plenty of catalogs of Coma cluster galaxies are available. The most frequently cited work is the catalog of \citet[GMP83]{GMP} listing magnitudes and colors, complete to $m_{b_{26.5}} = 20$, for $\sim$ 6700 galaxies centered on the Coma cluster core. However, it does not contain any redshift data or membership assignments. This information was later on added by \citet[MA08]{MA08} for a subsample of the GMP83 catalog, which is the principal data base of the present paper (s. Sect.~\ref{sec:MA08_sample}). Another important compilation is \citet[CD96]{CD96} who give 465 members with measured radial velocities, however being restricted, like most other catalogs, to relatively bright magnitudes \citep[e.g.,][]{Bej,Mob01,Lobo,Edw02}. Other catalogs go down to fainter magnitudes but are limited to a relatively small area \citep[e.g.,][]{Bernst,Trent}. There are also specific catalogs of dwarf galaxies in Coma \citep[e.g.,][]{SeckHar}. 

For our purposes, the catalog of choice has to (1) cover a sufficiently large area, (2) contain redshifts, (3) provide morphological information, (4) include faint galaxy populations. This is best met by the work of MA08 who give the morphology of all members and provide redshifts for more than $70\%$ of them, with a completeness limit of $M_B = -15$. Still, it 
does not cover the cluster outskirts, which is why we use the catalog of \citet[WLG11]{WL11} as a complementary data base (missing morphological information, though). 

\subsection{The MA08 catalog}
\label{sec:MA08_sample}

MA08 used the deep and wide field $B$ and $V$ imaging of the Coma cluster by \cite{Ad06} obtained with the CFH12K camera at the Canada-France-Hawaii Telescope (CFHT) to examine 1155 objects from the GMP83 catalog. Of these, 473 galaxies were assigned Coma cluster membership due to morphological appearance, apparent size, surface brightness, and redshift (for the detailed criteria see MA08). The data is complete down to $M_B = -15$ and extends to $M_B = -14.25$.\footnote{The MA08 catalog is free for download at\\ http://cdsweb.u-strasbg.fr/cgi-bin/qcat?J/A+A/490/923 \citep{MA08}.}. 

In the following sections a division between (bright) primary and (faint) secondary galaxies is applied (limit at $M_{B}=-19$). Of the 74 primaries a substantial fraction (29) is excluded due to their proximity to the cD galaxies and a general crowding in the central region. We formally exclude primaries in the innermost cluster region defined by an ellipse with minor axis 250 kpc (s. Sect.~\ref{sec:MA08_results}). The remaining set of 45 giants is further split into several subsamples of different morphological type. To keep the analysis statistically meaningful, the morphological classification given in MA08 had to be simplified. The classification bins used here, for primaries as well as secondaries, are given in Table~\ref{tab:P_morph}. The columns are: (1) The abbreviation used for the set (E/S0: early types, S: late types, dE: dwarfs, Irr: irregulars); (2) types included with the MA08 morphology assignments; (3) the number of primaries in each set; (4) the number of secondaries in each set (in parentheses: the number of secondaries with measured velocities). The numbers refer only to the region outside the central region indicated by the inner ellipse in Fig.\,1. 

\begin{table*}
  \caption[Morphology grouping of primary galaxies (brighter than $M_{B}=-19$)]{Morphology grouping of primary galaxies (brighter than $M_{B}=-19$) and secondary galaxies (fainter than $M_{B}=-19$)}
    \begin{tabular}{ccc>{\centering\arraybackslash}m{0.25\textwidth}}
    \hline\hline
    Morphology ID & MA08 Types included & Number of Primaries & Number of Secondaries (with measured $v$) \\
    (1) & (2) & (3) & (4)\\
    \hline
    \\
    E/S0  & cD, E/S0, S.0, E0-E5, Ep, SA0, SA0/a, SB0, SB0/a & 28 & 102 (99) \\
    \\
    S     & S.a, S.b, SAa-SBc & 17 & 66 (63) \\
    \\
    dE    & dE0-dE4, dS0, Ec & - & 184 (100)\\
    \\
    Irr   & Im, Sdm, Sm, Sm/dS0 & -& 47 (17)\\
    \hline
    \end{tabular}%
  \label{tab:P_morph}%
\end{table*}

The redshifts in the publicly available sources of the MA08 catalog are given only with three significant digits, for example as $z$ = 0.0231. This poses a problem for our velocity analysis in Sect.~\ref{sec:250kpc}, as the spectrum of velocity differences $\delta v$ becomes discrete instead of continuous. To get more accurate velocity data, we made a query in the NASA/IPAC Extragalactic Database (hereafter NED\footnote{http://ned.ipac.caltech.edu/}) and found velocities for 353 galaxies. Only these NED velocities are used here. Where more than one value was given, we chose the one with the smallest error (which is not necessarily the most reliable one). A Kolmogorov-Smirnov-Test and an Anderson-Darling-Test were performed to check that the velocity distribution is left unchanged, statistically, by the substitution of MA08 velocities with NED velocities.
The present work is based mainly on the sample described above, composed of the 473 members of the MA08 catalog, velocities taken from the NED, and the morphological classifications from Table~\ref{tab:P_morph}.  

\begin{figure*}[htbp]
\centering
\resizebox{15cm}{!}{\includegraphics[width=0.9\linewidth]{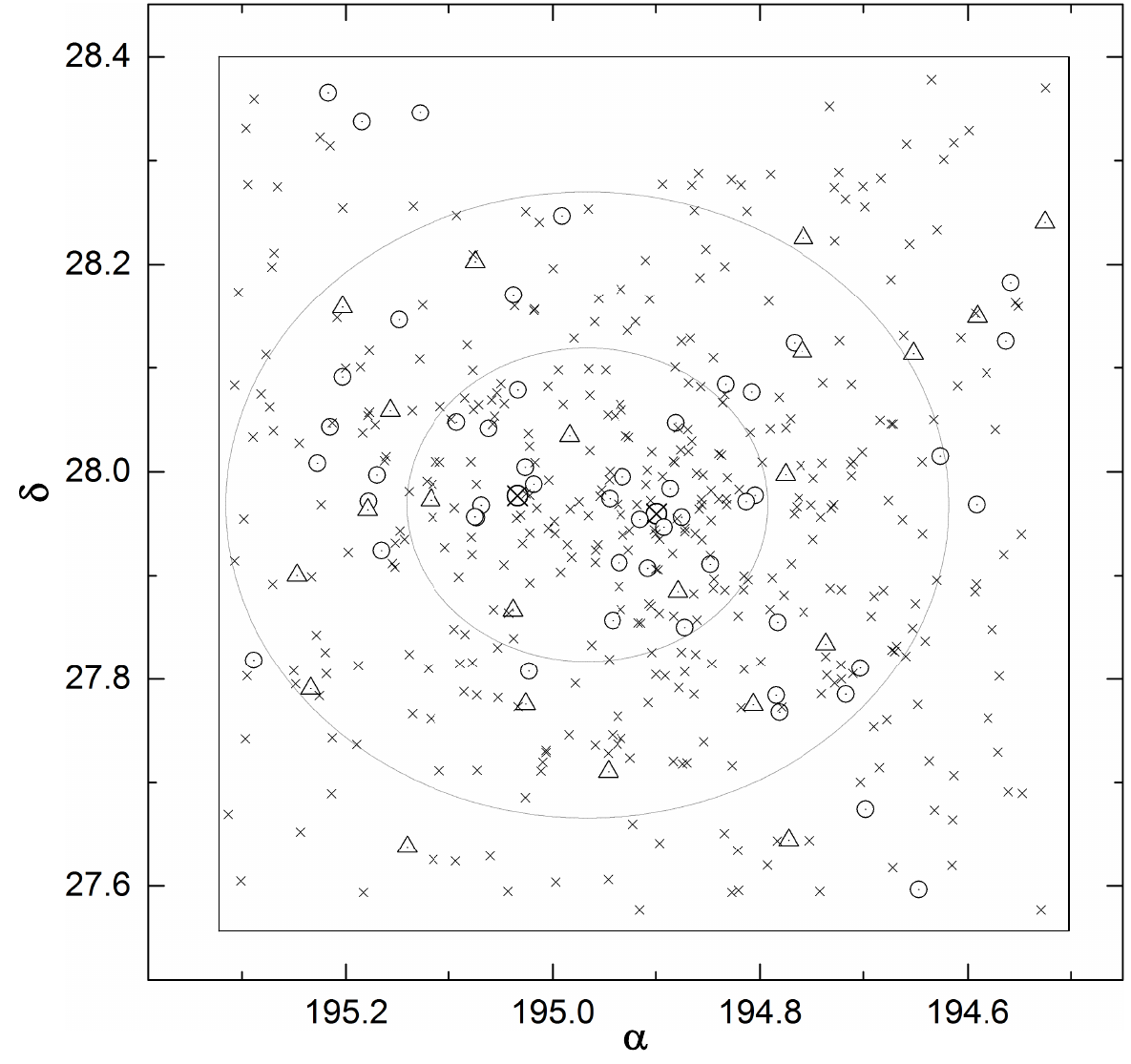}}
\caption[Inner and outer region for Coma primaries in the \citet{MA08} sample with morphological distinction]{Map of Coma cluster galaxies covered by MA08. Black crosses indicate galaxies fainter than $M_{B} = -19$ (the secondaries); circled dots: early-type primaries; dots in triangles: late-type primaries; bold circled crosses: NGC 4889 (left) and NGC 4874 (right). The inner (minor axis $b = \SI{250}{\kilo pc}$) and outer ($ b = \SI{500}{\kilo pc}$) gray ellipses separate the central, inner and outer regions considered in the analysis. Primaries within the innermost ellipse ($b = \SI{250}{\kilo pc}$) are excluded from the analysis. The rectangle indicates the area covered by the MA08 catalog which is relevant for density computations (s. Sect.~\ref{sec:MA08_sample}).}
\label{fig:Karte_Loc}
\end{figure*}

As the exact boundary coordinates of the CFHT field are not given in MA08, they are estimated here. The coordinates $x_{min}$, $x_{max}$, $y_{min}$, $y_{max}$ are approximated by the coordinates of the outermost galaxies covered (including non-members) plus a margin of $0.01^{\circ}$, such that $x_{min} =  \min\limits_i(\alpha_i)-\alpha_c-0.01$ or $y_{max} =  \max\limits_i(\delta_i)-\delta_c+0.01$ ($\alpha$ and $\delta$ for right ascension and declination in J2000, „c“ for central). So the resulting field has a coverage of $\sim 0.73 \times \SI{0.84}{\deg}^2$. In Fig.~\ref{fig:Karte_Loc} the positions of all galaxies in the MA08 catalog are shown. Primaries are divided morphologically, and the principal regions distinguished are indicated by the gray ellipses. 

\subsection{The WLG11 catalog}
\label{sec:WL11_sample}

\cite{WL11} studied the galaxy populations of several clusters including Coma and compared them to a semi-analytic model. To enlarge the sample of Coma members provided by MA08, these authors composed their own catalog of Coma cluster galaxies based on SDSS data, kindly made available to us by Dr. Thorsten Lisker, co-author of WL11. 

\begin{figure*}[htbp]
\centering
\resizebox{\hsize}{!}{\includegraphics[width=1\textwidth]{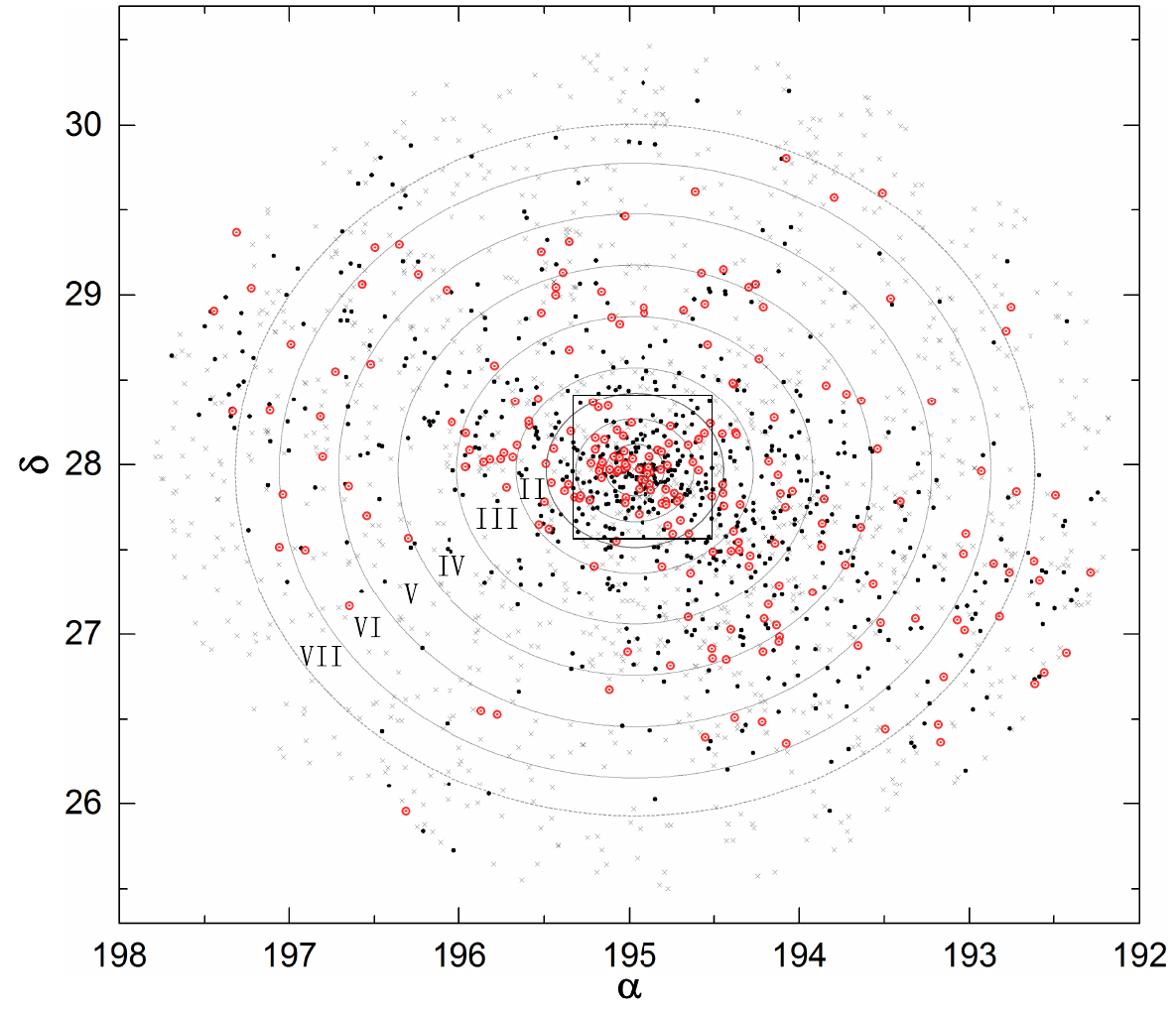}}
\caption[Map of Coma galaxies in the WLG11 sample]{Map of Coma cluster galaxies listed in WL11. Circled red dots: Primaries (brighter than $M_{R} = -20$); black dots: secondaries; small crosses: galaxies without redshifts which are not considered in this analysis. The subsamples, i.e. annular regions considered in Sect.~\ref{sec:WL11_Diff} are indicated by gray ellipses and Roman numbers (steps of $\sim \SI{0.5}{\mega pc}$). The black rectangle indicates the field covered by MA08 shown in Fig.~\ref{fig:Karte_Loc}.}
\label{fig:Karte_WL11}
\end{figure*}

The WL11 catalog is based on the SDSS DR7 \citep[$M_R \leq -16.7$]{SDSS} and covers a very large area (out to $\sim 2.5^\circ$ from the cluster center).
Cluster membership for 923 galaxies was determined spectroscopically. Another 383 galaxies were judged statistical members and 835 are statistical background (small crosses for both in Fig.~\ref{fig:Karte_WL11}). But only the 923 galaxies with measured redshifts are used here ($M_R \lesssim -17.2$). The sample contains 197 primaries and 695 secondaries (limit at $M_R = -20$) shown as red and black points in Fig.~\ref{fig:Karte_WL11}.

The WL11 catalog does not contain morphological information. But as it covers a much larger area (out to $\sim \SI{4.2}{\mega pc}$) than the MA08 catalog it is used here to study the outskirts of the cluster. Generally, it is difficult to determine a boundary of the cluster. \cite{LokM} for instance give a virial radius of $\sim \SI{2.8}{\mega pc}$. For practical reasons we defined a boundary simply by excluding all primaries with major axis distance from center $a_k$ larger than \SI{3.9}{\mega pc} (referring to the position shuffling method, s. Eq.~\ref{a-b-ratio2}). This limit is shown as outermost ellipse in Fig.~\ref{fig:Karte_WL11}. This guarantees that the density of secondaries within \SI{300}{\kilo pc} around each primary can be computed without applying any boundary corrections. For the MA08 data this is not the case, there boundary effects have to be corrected as described in Sect.~\ref{sec:density_comp}. Again the primaries close to the cD galaxies are not included (s. Sect.~\ref{sec:WL11_results}). 

\section{Method}
\label{sec:Method}

\cite{F92} split his sample of Virgo cluster galaxies \citep{VCC} into 83 primaries and 1157 secondaries (brighter or fainter than $M_B = -19$, respectively). He determined the surface density of secondaries around primaries by counting the number of secondaries within a step-wise growing distance from each primary divided by the area searched. To correct for projection effects Ferguson created pseudo-clusters by changing the positions of the primaries azimuthally around the cluster center (holding the radial distances fixed). This randomization was performed 40 times. Then he compared the mean surface densities ($\Sigma$) in the real data with the corresponding mean value from the pseudo-clusters ($\hat{\Sigma}$). Any excesses of the observed over the expected numbers (surpassing one standard deviation) were finally interpreted as indication for the presence of bound companions; the statistical significance of the excesses was assessed by a Kolmogorov-Smirnov-Test. This kind of procedure was repeated with a subsample of galaxies that have measured redshifts (79 primaries and 290 secondaries). Small velocity differences ($\lesssim \SI{250}{\kilo m.s^{-1}}$) between primaries and secondaries were taken as additional evidence for their gravitational binding. A further refinement of the analysis was done with respect to primary position in the cluster, primary morphology and luminosity. F92 also introduced an interaction parameter to compare a galaxy's strongest local interaction to  its interaction with the rest of the cluster. 

This is the basic procedure adopted for the present Coma cluster analysis, with only slight deviations which are discussed in the following. In Sect.~\ref{sec:Method_Density}-\ref{sec:MC_prop} the generation of the pseudo-clusters and the computation of $\Sigma$ and $\hat{\Sigma}$ are explained. The interaction parameter method is described in Sect.~\ref{sec:IP_Shuffling}. 	

\subsection{Monte Carlo methods and companion density $\Sigma(r)$}
\label{sec:Method_Density}

\subsubsection{Generation of uniformly distributed primary positions}
\label{sec:position_method}
The generation of pseudo-clusters is done by randomly assigning new positions to the primaries uniformly along an ellipse around the center of the Coma cluster ($\alpha_c = \SI{194.9668}{\degree},\delta_c = +\SI{27.9680}{\degree}$, WL11). For this purpose a coordinate system $(x,y)$ is introduced with the cluster center as its origin.

F92 shuffled the Virgo primaries azimuthally, simply in want of a clear symmetric form of the cluster. However, the Coma cluster is more regular and exhibits a fairly well defined elliptical shape \citep[e.g.,][]{BiEps05,CaMet}. It is therefore more adequate for Coma to randomize the primary positions uniformly along ellipses instead of circles, based on the following equation.

\begin{equation}
\label{a-b-ratio2}
\frac{x_k^2}{a_k^2}+\frac{y_k^2}{b_k^2} = 1 \quad
\Leftrightarrow \quad \sqrt{x_k^2 (1-e)^2 + y_k^2} = |b_k|\,\,\,,
\end{equation}
with $(x_k,y_k)$ as position of the $k$-th primary. Substituting $b_k$ back into the left formula leads to $a_k$. 
The ellipticity $e = 1 - \frac{b}{a}$ 
($a,b$ are the semi-major and semi-minor axis lengths) is estimated by different ways (e.g., from the standard deviation in x- and y-coordinates of the galaxies). For Coma we find $e \approx 0.13$.

The elliptical shuffling method is non-trivial. A simple drawing of a random azimuthal angle $\phi_{ran}$ (uniformly distributed) in polar coordinates would lead to a non-uniform distribution of random positions on an ellipse. 
The same holds true for points uniformly distributed along a circle which are linearly transformed to an ellipse.
Thus, an acceptance-rejection algorithm is performed using two probability density functions (pdfs) $g(t),f(t)$ such that ${C \cdot g(t) \geq f(t)} ~ \forall t$ and for some $C \geq 1$. We construct $f(t)$ from the two-dimensional line element ${ds= a\sqrt{1  - \epsilon^2\cos^2(t)}}$ 
and the perimeter $\mathcal{L}$ of the ellipse:

\begin{equation}
\label{eq:pdf}
f(t) = \frac{ds(t)}{\mathcal{L}} = \frac{\sqrt{1  - \epsilon^2\cos^2(t)}}{\int_0^{2\pi} \sqrt{1  - \epsilon^2\cos^2(t)} dt}\,\,\,,
\end{equation}
where $\epsilon$ is the eccentircity; $\epsilon^2 = 1-b^2/a^2 = 1-(1-e)^2$.
We use ${g(t) = \mathcal{U}[0,2\pi]}$ and ${C = \frac{2\pi a}{\mathcal{L}}}$. In this way the algorithm produces $\phi_{ran} \sim f(t)$. As a last restriction the positions drawn at random must fall into the CFHT frame of the MA08 catalog. This is implemented directly into the acceptance-rejection algorithm. For the WL11 primaries no problems occur since all major axes $a_k$ are smaller than \SI{3.9}{\mega pc}. 

\subsubsection{Velocity shuffling}
\label{sec:V-Shuffle}
Another possibility to create pseudo-clusters is the shuffling of primary velocities. To keep the same overall velocity distribution the new radial velocities are not just randomly generated numbers. Instead all the primary velocities of the sample are redistributed randomly to a new primary. 
This is achieved by implementing the Fisher-Yates shuffle \citep[also known as Knuth shuffle,][p.542]{TrendsNet} which produces permutations ${\pmb{v_{\tau}} = (v_{\tau_1}, v_{\tau_2},...,v_{\tau_N})}$ of the primary velocity vector $\pmb{v}$ with equal probabilities for each permutation $\tau$. With this algorithm also primaries which are placed closer to the cluster center than \SI{250}{\kilo pc} can be included. Positions stay unchanged in these pseudo-clusters. 

\subsection{Computation of $\Sigma(r)$}
\label{sec:density_comp}

To compute the secondary density around primaries (hereafter simply referred to as density) a circle with radius $r$ is chosen around a particular primary galaxy and all galaxies within this area are counted and divided by the area. This is done for a series of growing radius, that is the separation $r$ is grouped into bins $[0,r_i]$. Thus the area searched is always a circle and not an annulus, and the densities distributions plotted for the binned radii $r_i$ are cumulative, not differential. For differential distributions the numbers of galaxies are too small for statistical purposes. Computations in Sect.~\ref{sec:RESULTS} are done with

\begin{equation}
\label{eq:dens0}
   \begin{aligned}
   \Sigma_i &= \Sigma(r_i)= N_i/A_i\,\,,\\
   A_i &= r_i^2\pi\,\,.
   \end{aligned}
   \end{equation}

$N_i$ is the number of secondary galaxies within $A_i$ (area of the $i$-th circle). As Coma does not subtend a large solid angle, all galaxies are assumed to lie on the same plane tangential to the sky and euclidean distances $r$ are calculated.
For primaries closer than \SI{300}{\kilo pc} to a boundary, the density calculated by Eq.~\ref{eq:dens0} has to be corrected by taking into account the effective search area falling inside the boundaries $(A_{eff})$:
 
\begin{equation}
\Sigma_{i,corr} = \frac{\Sigma_i \cdot A_i}{A_{eff}}\,\,\,.
\end{equation}

\subsection{Properties of MC results}
\label{sec:MC_prop}

To compare the densities $\Sigma(r_i)$ thus computed for the real cluster averaged over all primaries 
with the expected densities from pseudo-clusters, we had to calculate the corresponding mean density $\Sigma^j(r_i)$ for each single Monte Carlo run ($j = 1,2,...,n$) and finally 
the mean density $\hat{\Sigma}(r_i)$ and standard deviation $\hat{\sigma}_i$ for the whole set of $n$ pseudo-clusters.

The Monte Carlo (MC) results were tested for stability in different ways. Numerous experiments with various $n \geq 50$ of MC runs were executed and the results did not show any significant changes. Thus, we can assume to have a stable situation and reliable results by choosing $n = 1000$ for the position and velocity randomizations. Different random number generators (RNG) were tested, again without noticeable differences. Our MC runs were done with the standard and very rapid `Mersenne Twister' RNG algorithm which provides extremely uniformly distributed random numbers. Using different seeds, that is with the current time stamp, with zero or prime numbers, yielded stable results. The large period of $2^{19937}-1$ ($\approx 4 \cdot 10^{6001}$) guarantees the avoidance of correlated random numbers. 

\subsection{Interaction parameter and direct position shuffling method}
\label{sec:IP_Shuffling}

\subsubsection{The interaction parameter method}
\label{sec:IP_theory}
F92 used a second approach to look for bound companions which avoids the need to distinguish between primaries and secondaries. The basic idea is to define a nearest neighbor of each galaxy not in terms of spatial proximity but of gravitational influence. F92 introduced an interaction parameter (IP) to identify each galaxy’s neighbor with the largest gravitational pull $\mathcal{M}/r^2$ ($r$ describing the projected separation between the galaxies). The mass $\mathcal{M}$ of a galaxy is notoriously difficult to know but can be sufficiently well represented by its luminosity $L$. We apply this method here as well. For the MA08 sample we employ $L$ in units of blue solar luminosity $L_{\sun,B}$. 
For galaxies of the WL11 catalog $L_R$ and $M_{R}$ are used instead. 

The principal IP used here is the same as in F92 but defined directly by $L$ instead of $M$:

\begin{equation}
I_{Ferg,l} = \frac{\underset{k}{max} \left( L_{k}/r^2_{lk} \right)}{\sum\limits_{k \neq k_{max}} \left( L_{k}/r^2_{lk} \right) }\,\,\,.
\label{eq:IP_Ferg_theory}
\end{equation}

For the $l$-th galaxy this IP compares the interaction with its nearest neighbor (indexed as $k_{max}$) to that with the rest of the cluster. In Sect.~\ref{sec:IP_2_05} galaxies which have $I_{Ferg} > 2$ (strongest interaction with one galaxy twice as high than with all others) will be considered as bound (F92). This is only valid in a statistical way and may be wrong for a particular galaxy. \citeauthor{F92}'s IP clearly does not take into account the cancellation of gravitational forces by different neighbors nor the three dimensional structure of the cluster. Also, the analysis of $\Sigma$ is always performed up to separations of \SI{300}{\kilo pc}, while according to the definition of $I_{Ferg}$ in Eq.~\eqref{eq:IP_Ferg_theory} the ‚nearest neighbor’ of a galaxy might have a larger separation. 

We explore alternative IPs that take more than one influential neighbor into account. First, a substructure IP which compares the gravitational interaction for a galaxy with all its neighboring galaxies within \SI{300}{\kilo pc} ($k'$) to the interaction with the rest of the cluster ($k''$, separations $r_{lk''} > \SI{300}{\kilo pc}$): 

\begin{equation}
\label{eq:IP_substructure_theory}
I_{sub,l} = \frac{\sum\limits_{k'} \left( L_{k'}/r^2_{lk'} \right)}{\sum\limits_{k''} \left( L_{k''}/r^2_{lk''} \right) }\,\,\,.
\end{equation}

This IP should show a significantly higher fraction of galaxies with $I_{sub} > 2$ than \citeauthor{F92}'s IP. 

As a second variant we define a group IP by comparing the three most strongly interacting galaxies (if those galaxies are closer than \SI{300}{\kilo pc}, $k'$) with the interaction by the rest of the cluster:

\begin{equation}
\label{eq:IP_group_theory}
I_{group,l} = \frac{\underset{k'}{max^*} \left( L_{k'}/r^2_{lk'} \right) + \underset{k' \neq k_{max}}{max^*} \left( L_{k'}/r^2_{lk'} \right)+\underset{\substack{k' \neq k_{max}\\ k' \neq k_{max,2}}}{max^*} \left( L_{k'}/r^2_{lk'} \right)}{\sum\limits_{\substack{k \neq k_{max}\\k \neq k_{max,2}\\k \neq k_{max,3}}}\left( L_{k}/r^2_{lk} \right) }\,\,\,,
\end{equation} 
where "$*$" denotes that those maxima are only computed if the corresponding galaxies are located closer than \SI{300}{\kilo pc} to the $l$-th galaxy.

This IP should be sensitive to the presence of subgroups within the cluster. We assume that if such subgroups exist, $I_{group}$ should show approximately the same values for the galaxies contained in a group. Therefore the number of galaxies with $I_{group} > 2$ should range somewhere between the number of $I_{Ferg}$ and $I_{sub}$.

As hitherto only gravitational interaction is considered, a last IP is introduced which is sensitive to tidal interaction. It is almost the same IP as the one in \citet[hereafter Ka05]{Ka05}: 

\begin{equation}
\label{eq:IP_Ka05_theory}
I_{Ka05,l} = \underset{k}{max}\left[ log(L_{B,k}/r^3_{lk}) \right]\,\,\,.
\end{equation}
The only difference is that in Ka05 a constant $C$ was added which allows to identify isolated galaxies ($I_{Ka05,l} < 0$). 

\subsubsection{Direct position shuffling method}
\label{sec:direct_pos_Meth}
The pseudo-clusters to calculate IPs which then can be compared with the real IPs based on the MA08 and WL11 catalogs are created by a direct position shuffling. In this method the positions of two randomly chosen galaxies $k$ and $l$ are swapped directly.
As in F92 this exchange is performed $\mathcal{N}^2$ times where $\mathcal{N}$ is the total number of galaxies in the sample. In this way it can happen that a galaxy changes its position several times in one run, while another may stay in place. 


\section{Results}
\label{sec:RESULTS}
\subsection{Results for the $\Sigma(r)$-analyses of the MA08 catalog} 
\label{sec:MA08_results}

The settings for the MC simulations analyzed in this section are given below. They are valid throughout this paper, unless mentioned otherwise.

\begin{itemize}
\item Primary-secondary discrimination at $M_{B}=-19$ (74 giants and 399 possible companions at the outset)
\item Primaries have to lie outside an ellipse with minor axis 250 kpc centered on the cluster center (45 primaries left)
\item Distance to Coma: \SI{100}{\mega pc}, $H_0=\SI{70}{\km.s^{-1}.\mega pc^{-1}}$, $\Omega_{\Lambda}=0.7$, $\Omega_m=0.3$, Distance modulus $\mu=35$ and therefore a scale of $\SI{0.46}{\kilo pc.\arcsec^{-1}}$ \citep{A09} 
\item To calculate the absolute magnitudes, a small correction of $0.25$ is added to $\mu$ leading to a value of $\mu'=35.25$ (MA08).
\item The cluster center is defined as in WL11 at $\alpha_c = \SI{194.9668}{\degree}$ and $\delta_c = +\SI{27.9680}{\degree}$.
\item $\Sigma(r)$ is computed up to \SI{300}{\kilo pc} for bins $[0,r_i]$ with $r_{i+1} = r_i + \SI{20}{\kilo pc}$ and $r_1 = \SI{20}{\kilo pc}$.
\item Random number generator (RNG): Mersenne Twister seeded with the current time stamp
\item $n=1000$ Monte Carlo runs
\end{itemize}

In order to avoid any influence on the companion density by primaries lying in the crowded central region around the two cD galaxies, the primary set underlies a restriction: As the position shuffling is performed elliptically, the computed minor axis $b_k$ (s. Eq.~\eqref{a-b-ratio2}) for the randomization of any primary used has to be larger than \SI{250}{\kilo pc}. This prevents the random placement of a primary within an ellipse with $b = \SI{250}{\kilo pc}$ around the cluster center, which is located midway between NGC 4874 and NGC 4889. We note that the point of this exclusion is not only to avoid the crowded region as such; the problem is that the azimuthal or elliptical randomization that close to the center produces overlapping sets of secondaries. F92 chose a radial distance of \SI{200}{\kilo pc} for the center of the Virgo cluster. This distance is slightly extended here due to the presence of substructure around the cD galaxies \citep[e.g.,][]{N01}. $\hat{\Sigma}(r)$ of the remaining 45 primaries is stable even if the ellipticity is slightly changed, which also supports the appropriateness of the chosen value of \SI{250}{\kilo pc}.

In this section, the results of the MC simulations with the settings above are presented. There are mainly two kinds of outcomes: Firstly, the surface density of companions around primaries and secondly, the distribution of separations between primaries and secondaries. Both sets are further split into several subsets in Sect.~\ref{sec:250kpc_PrimMorph}.

As an additional criterion to search for physically bound companions, the velocity differences $\delta v$ between primaries and secondaries can be used. Results found in this way have a stronger weight, especially because the observed field is not very large and the areas searched for satellites are quite often overlapping. As mentioned in Sect.~\ref{sec:MA08_sample} we take all velocities from the NED. Measured velocities are available for all primaries but only for  part of the secondaries (see Table~\ref{tab:P_morph}). 

F92 grouped $\Sigma$ into four bins of $\delta v$ from \SI{125}{\kilo m.s^{-1}} up to \SI{500}{\kilo m.s^{-1}}. Here we use two bins of velocity differences only: $\SI{0}{\kilo m.s^{-1}} \leq \delta v \leq \SI{250}{\kilo m.s^{-1}}$ and $\SI{250}{\kilo m.s^{-1}} < \delta v \leq \SI{500}{\kilo m.s^{-1}}$. Over-densities for velocity differences in the first bin are a strong indication of physical interaction, that is gravitational binding of primary and secondary galaxies. When $\delta v$ lies in the second bin, we expect a massive primary or a whole subgroup able to hold its satellites over a larger separation. In `calmer' outer regions we generally expect a higher fraction of bound satellites. 

Two methods can be employed for galaxies with redshifts. The first method is the usual elliptical position randomization of the 45 primaries, where $\Sigma(r)$ and $\hat{\Sigma}(r)$ are determined only for the 279 secondaries with a velocity. In the second method the primary velocities are shuffled to create a pseudo-cluster (s. Sect.~\ref{sec:V-Shuffle}, also referred to as $v$-shuffling). Evidence is only taken as acceptable if both methods produce similar results. For the central primaries (minor axis smaller than \SI{250}{\kilo pc}) $\Sigma(r)$ can be analyzed only with the second randomization method. This set contains additional 24 early-type (including both cD) and 4 late-type galaxies. 

\subsubsection{Complete set of primaries and secondaries}
\label{sec:250kpc}

\begin{figure*}[hbt]
\centering
\resizebox{15cm}{!}{\includegraphics[width=0.85\textwidth]{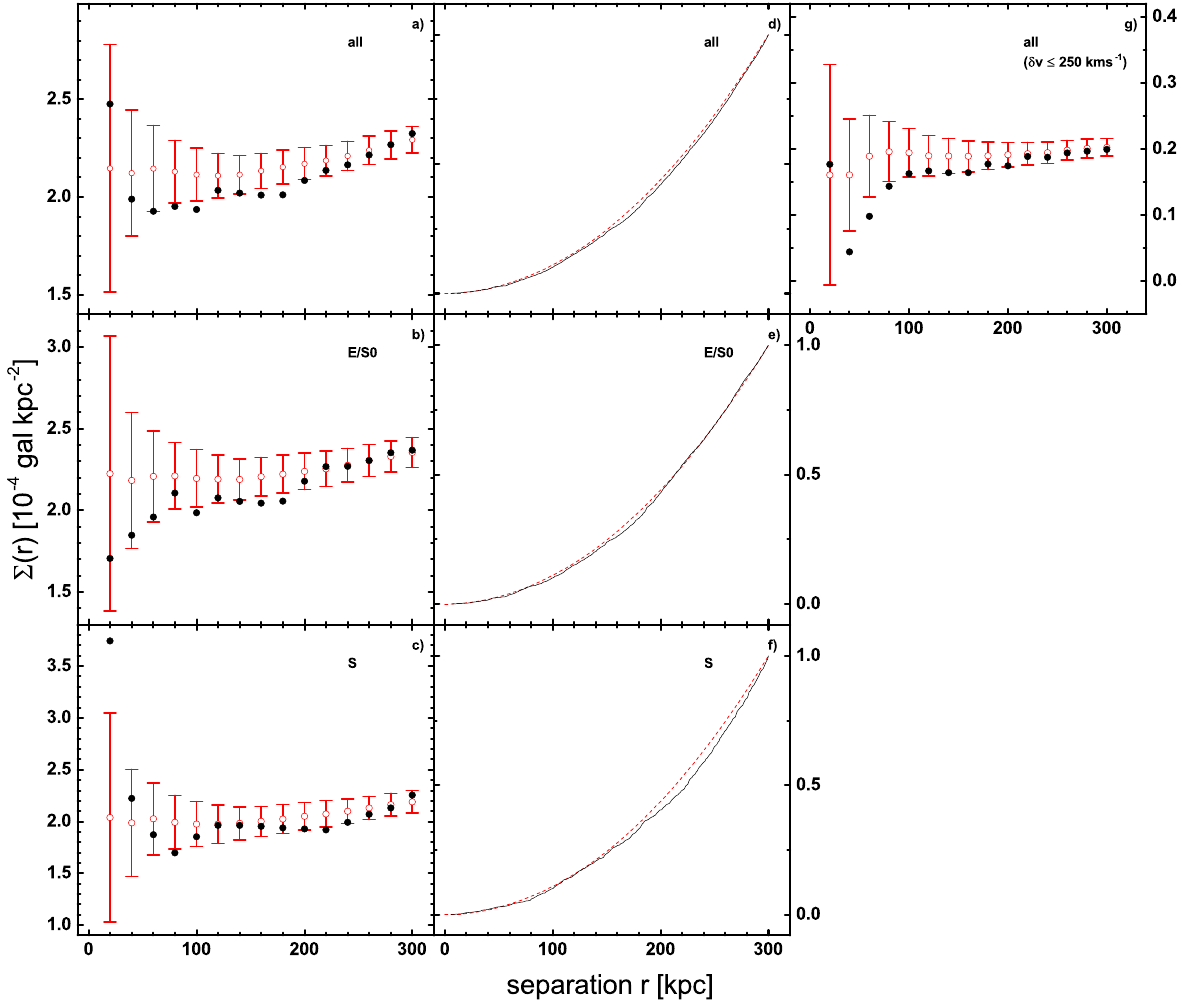}}
\caption{a)-c): Mean density of secondaries around primaries depending on their separation r. It should be noted that the $r_i$ do not indicate annular bins. $\Sigma(r_i)$ represents the number of secondaries within a circle of radius $r_i$ (s. Sect.~\ref{sec:density_comp}). Black dots indicate the values for the MA08 set, the open circles and error bars (in red) show the mean and standard deviation of 1000 pseudo-clusters — a): whole sample, b): early-type primaries only, c): late-type primaries only. 
\newline
d)-f): Cumulative distribution of primary-secondary separations $r$ for the MA08 data (black) and 1000 pseudo-clusters (dashed, red) — d): whole sample, e): early-type primaries only, f): late-type primaries only.
\newline
g): The same as a) but restricted to galaxies with velocity difference smaller than 250 km $s^{-1}$.}
\label{fig:8Var}
\end{figure*}

In Fig.~\ref{fig:8Var}a, top left, we show the surface density of all secondaries around all 45 primaries. The open circles with error bars indicate the mean density $\hat{\Sigma}_i$ and standard deviation $\hat{\sigma}_i$ of 1000 MC runs. The black dots give the mean densities $\Sigma_i$ of the data from MA08. This type of figure is the  basic tool of our analysis.

The effects sought for should fulfil one of the following conditions: 
1) The mean MA08 density $\Sigma_i$ lies outside the error bars of the pseudo-clusters at least for three bins in a row (on the same side, significance at $1\hat{\sigma}$ for $r_i, r_{i+1}, r_{i+2}$), or 2) the observed density deviates from the model at the $2\hat{\sigma}$-level or more for at least one bin. 

As can be seen, there is no such effect with respect to either specification. Instead of the expected over-densities there are rather under-densities. Toward the largest separation the observed and model densities match, as required, but for smaller separations the observed densities are systematically too low, which we ascribe to a combination of different biases stemming from local deviations from the assumed cluster ellipticity etc. For further analysis, the sets are differentiated in the following subsection. 

The same is found by setting an upper limit for the velocity separation, $\delta v \leq \SI{250}{\kilo m.s^{-1}}$, as shown in Fig.~\ref{fig:8Var}g, upper right. The effect is even slightly accentuated by the apparently missing companions for small primary-secondary separations at $r_i = 40 - \SI{80}{\kilo pc}$.

Another way to compare real-cluster with pseudo-cluster data is to calculate the `cumulative distribution functions' (CDF) for the separation of primaries and secondaries for both and to analyze the difference by a Two-Sample-Kolmogorov-Smirnov-Test (KS-Test).
There is one fundamental difference to the analysis with densities $\Sigma(r)$, namely one must assume that the borders of the field do not have any influence on the distribution (F92), as it is no longer possible to simply divide the numbers by the area searched. 
 
The null hypothesis stating that the CDF of the primary-secondary separations of the real cluster and the pseudo clusters are drawn from the same unknown background distribution is rejected if the p-value of the KS test is smaller than a significance level of $5\%$. As the KS test is rather conservative \citep{EngCou}, for p-values close to the $5\%$-significance level a Anderson-Darling-Test (AD test) is performed in addition. While the KS-Test is sensitive with respect to global differences, the AD test is more powerful to discover differences near $0$ and $\SI{300}{\kilo pc}$ where the CDF converges to 0 and 1 respectively \citep[e.g.,][]{EngCou}. Furthermore, the AD-Test needs less data to reach sufficient statistical power. However, for very small numbers (e.g.,less than $\sim 10$ primaries) this test is performed neither. 

The CDF for the whole sample is shown in Fig.~\ref{fig:8Var}d, top middle panel. The KS test does not reject the null hypothesis but the additionally performed AD test does point to a difference in the distributions (p-values are $p_{KS} = 0.080$ and $p_{AD} = 0.046$, respectively, see Fig.~\ref{fig:8Var}d). However, if only galaxies with measured velocities are used for the tests, there is no evidence for different background distributions left. 

\subsubsection{Dependence on type, $\delta v$, location in cluster, and luminosity}
\label{sec:250kpc_PrimMorph}

The 45 primaries encompass 28 early-type (E/S0) galaxies and 17 late-type (spiral) galaxies (see Table~\ref{tab:P_morph}). The secondary densities for these two morphological subgroups are shown in Fig.~\ref{fig:8Var}b,c. The under-densities that appear for the whole sample in Fig.~\ref{fig:8Var}a are clearly present also for the early-type primary sample (Fig.~\ref{fig:8Var}b) but not for spiral primaries (Fig.~\ref{fig:8Var}c), that is the effect is due to early-type primaries. When the velocities are restricted as for the whole sample in Fig.~\ref{fig:8Var}g, there is no such effect seen in either subset (not shown here), casting doubt on the reality or relevance of the under-densities. 
More confusing still are the CDF’s for the two morphological subgroups, shown in Fig.~\ref{fig:8Var}e,f. Countering the expectation from the density results, here it is the E/S0 sample that is indifferent ($p_{KS} = 0.208$), while the spiral primary sample exhibits significant under-densities in the separation range $r \sim 150 - \SI{260}{\kilo pc}$, with a strong signal ($p_{KS} = 0.005$ and $p_{AD} = 0.004$, — and if restricted to secondaries with velocities $p_{KS} = 0.031$ and $p_{AD} = 0.020$). This contradiction calls for further differentiation of the sample.

First we look for any dependence on the location of the primaries in the cluster by splitting the primary sample into an inner set of 33 primaries located within an ellipse of minor axis $b$ restricted to $\SI{250}{\kilo pc} < b < \SI{500}{\kilo pc}$ and an outer set of 12 primaries with $b \geq \SI{500}{\kilo pc}$ (this division is indicated in Fig.~\ref{fig:Karte_Loc}; remember that the innermost region $b < \SI{250}{\kilo pc}$ is avoided because of confusion). However, this splitting alone does not reveal any new results. So the positional criterion is combined with morphology, yielding four subsamples encompassing  
21 inner early-types, 12 inner late-types, 7 outer early-types, and 5 outer late-types. By restricting the sets additionally to secondaries with velocity data, some effects do appear, but almost all of them concern the outer samples where statistical testing is impossible due to the small numbers involved. We find only a significant signal for the inner late-type set ($p_{KS} = 0.023$), again for under-densities, that is apparently missing companions.

\begin{figure}[htbp]
\resizebox{\hsize}{!}{\includegraphics[width=1\linewidth]{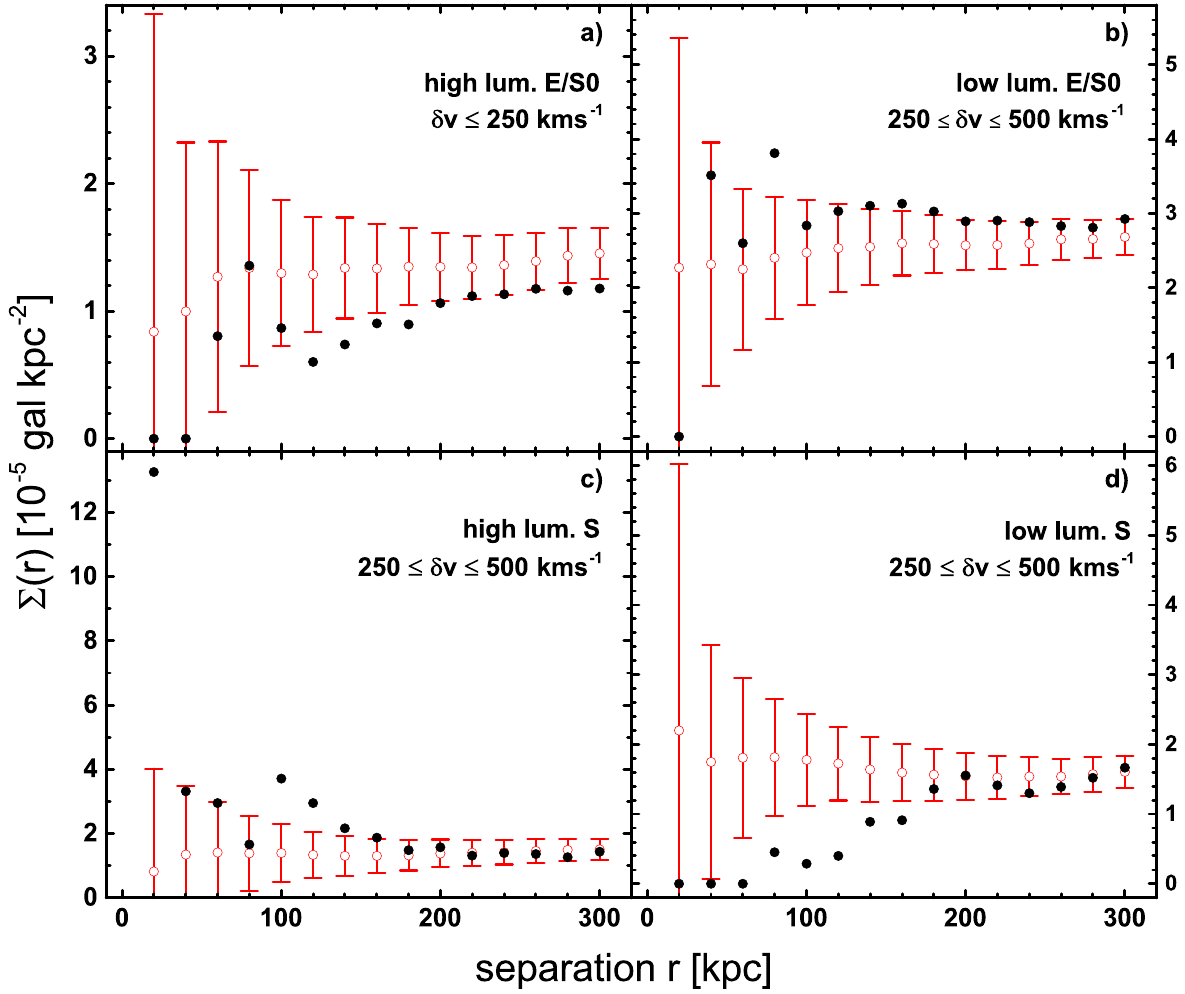}}
\caption{Mean density of secondaries around different subsets of primaries. The set used, along with the $\delta v$ restriction, is indicated in each panel. Full black dots represent the data; colored open circles and error bars the mean and standard deviation of 1000 pseudo-clusters (created by shuffling primary positions).}
\label{fig:Lum_E_S}
\end{figure}

Next, we try a differentiation with respect to primary luminosity by dividing the whole into 17 high-luminosity ($M_{B} < -20$) and 28 low-luminosity primaries ($-19 \geq M_{B} \geq -20$). The only significant result is a $2\hat{\sigma}$-excess for $r_i = \SI{20}{\kilo pc}$ and large $\delta v$ in the high-luminosity sample. More revealing is the combination of luminosity and morphology, while nothing significant is found for the combination of luminosity and location. Adding the $\delta v$ criterion finally gives the following effects, see Fig.~\ref{fig:Lum_E_S}:

\begin{itemize}
\item Excess of possible companions around high-luminosity late-types (s. Fig.~\ref{fig:Lum_E_S}c)
\item A lack of secondaries around high-luminosity early-types and low-luminosity late-types (s. Fig.~\ref{fig:Lum_E_S}a and d)
\item A small excess around low-luminosity E/S0 giants (s. Fig.~\ref{fig:Lum_E_S}b)
\end{itemize}
For the low-luminosity late-type set we find statistical confirmation of under-densities by the KS test ($p_{KS} = 0.040$). 

\begin{figure}[htbp]
\resizebox{\hsize}{!}{\includegraphics[width=1\linewidth]{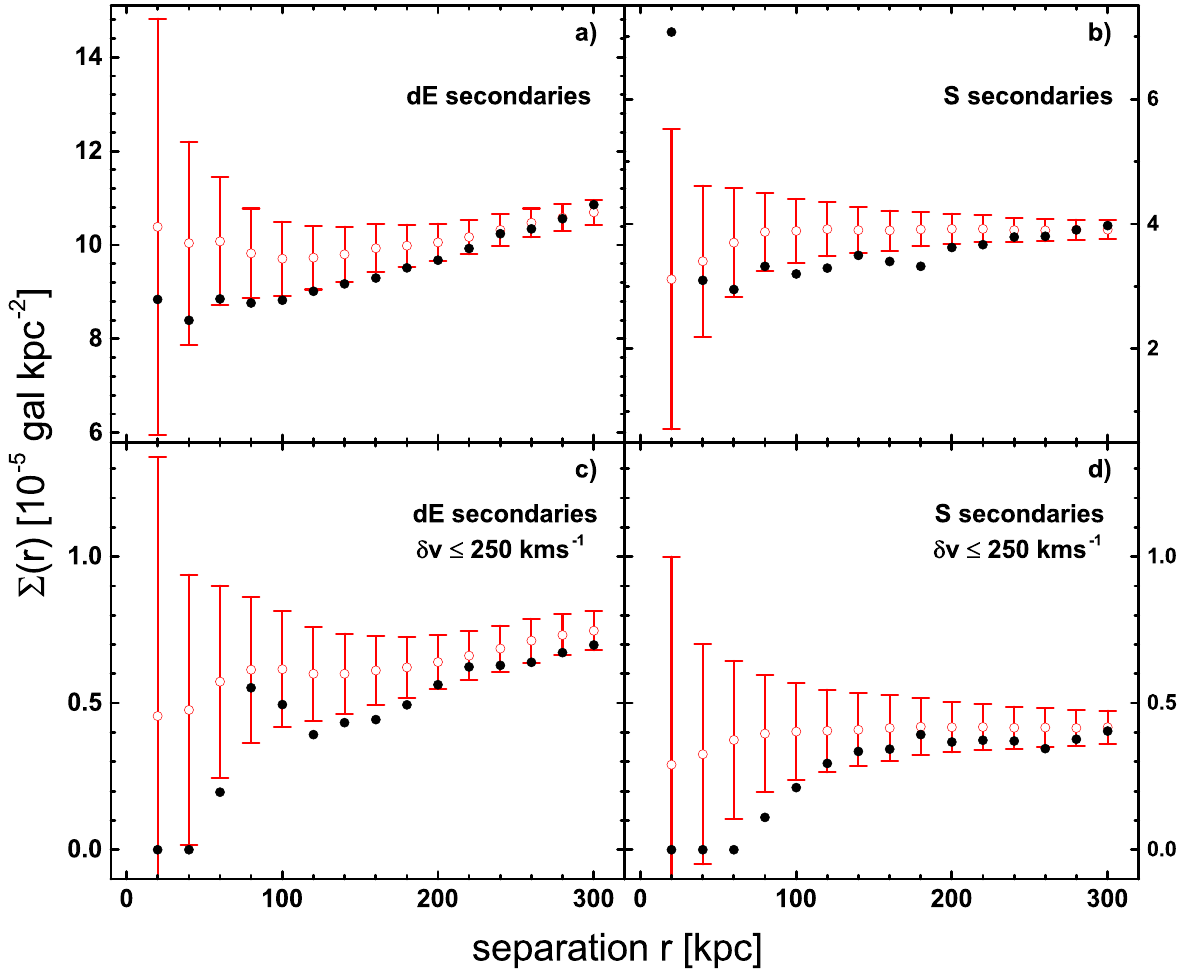}}
\caption{Mean density of dwarf elliptical (dE, left panels) and spiral (S, right panels) secondaries around all primaries. In the lower panels the same is shown for primary-secondary pairs with a small velocity difference $\delta v < 250$ km s$^{-1}$ only. Full black dots represent the data; colored open circles and error bars the mean and standard deviation of 1000 pseudo-clusters.}
\label{fig:Sec_Morph}
\end{figure}

\begin{figure}[htbp]
\centering
{\includegraphics[width=0.6\linewidth]{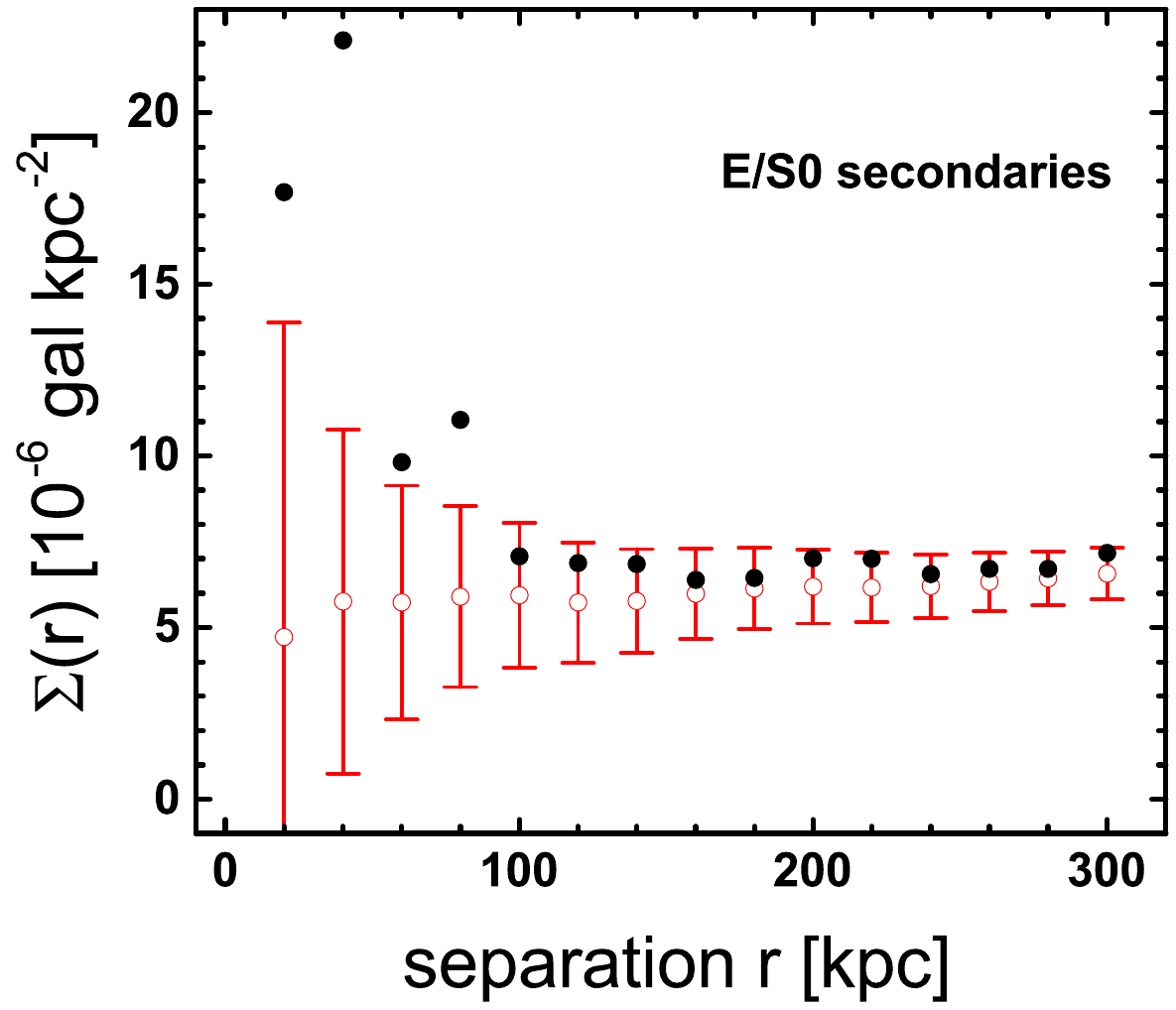}}
\caption{Mean density of E/S0 secondaries around all primaries with large velocity differences (\mbox{$250 \leq \delta v \leq \SI{500}{\kilo m.s^{-1}}$}). Full black dots represent the data; colored open circles and error bars the mean and standard deviation of 1000 pseudo-clusters.}
\label{fig:Sec_nurE}
\end{figure}

In the following the morphology of the 399 secondaries is considered. The secondaries are divided into four groups, as defined in Table~\ref{tab:P_morph}: 102 early-type (E/S0) galaxies, 66 late-type (S) galaxies, 184 early-type dwarfs (dE) and 47 irregulars (Irr). The density of each subsample of secondaries is first computed around all primaries and then also around the primary subsamples (inner/outer and early/late). The main results are shown in Figs.~\ref{fig:Sec_Morph} and \ref{fig:Sec_nurE}. 

Overall, secondaries of any morphological type tend to be depleted out to $\sim \SI{200}{\kilo pc}$. This is exemplified for dwarf ellipticals and spirals in Fig.~\ref{fig:Sec_Morph} (however, we note the excess for the very first separation bin of spiral secondaries). This holds true even when the velocity difference is restricted to $\delta v \leq \SI{250}{\kilo m.s^{-1}}$, where companions are expected most. In contrast, and surprisingly, when larger velocity difference are considered
($250 \leq \delta v \leq \SI{500}{\kilo m.s^{-1}}$, that is when stronger binding forces are required, there is a strong hint for the presence of E/S0 companions at small separations, as seen in Fig.~\ref{fig:Sec_nurE}. Unfortunately, this remains unconfirmed by KS or AD testing.

As a last breakdown of the sample also the secondary luminosity is regarded. The secondaries are divided into five groups: 62 galaxies with $-19 < M_{B} \leq -18$; 74 (only 72 with measured $v$) with $-18 < M_{B} \leq -17$; 86 (74) with $-17 < M_{B} \leq -16$; 98 (47) with $-16 < M_{B} \leq -15$; and 79 (24) with $M_{B} > -15$. In the last bin probably more galaxies exist, but the MA08 catalog is only complete down to the limit of $-15$. 

\begin{figure*}[htb]
\centering
\resizebox{15cm}{!}{\includegraphics[width=0.85\textwidth]{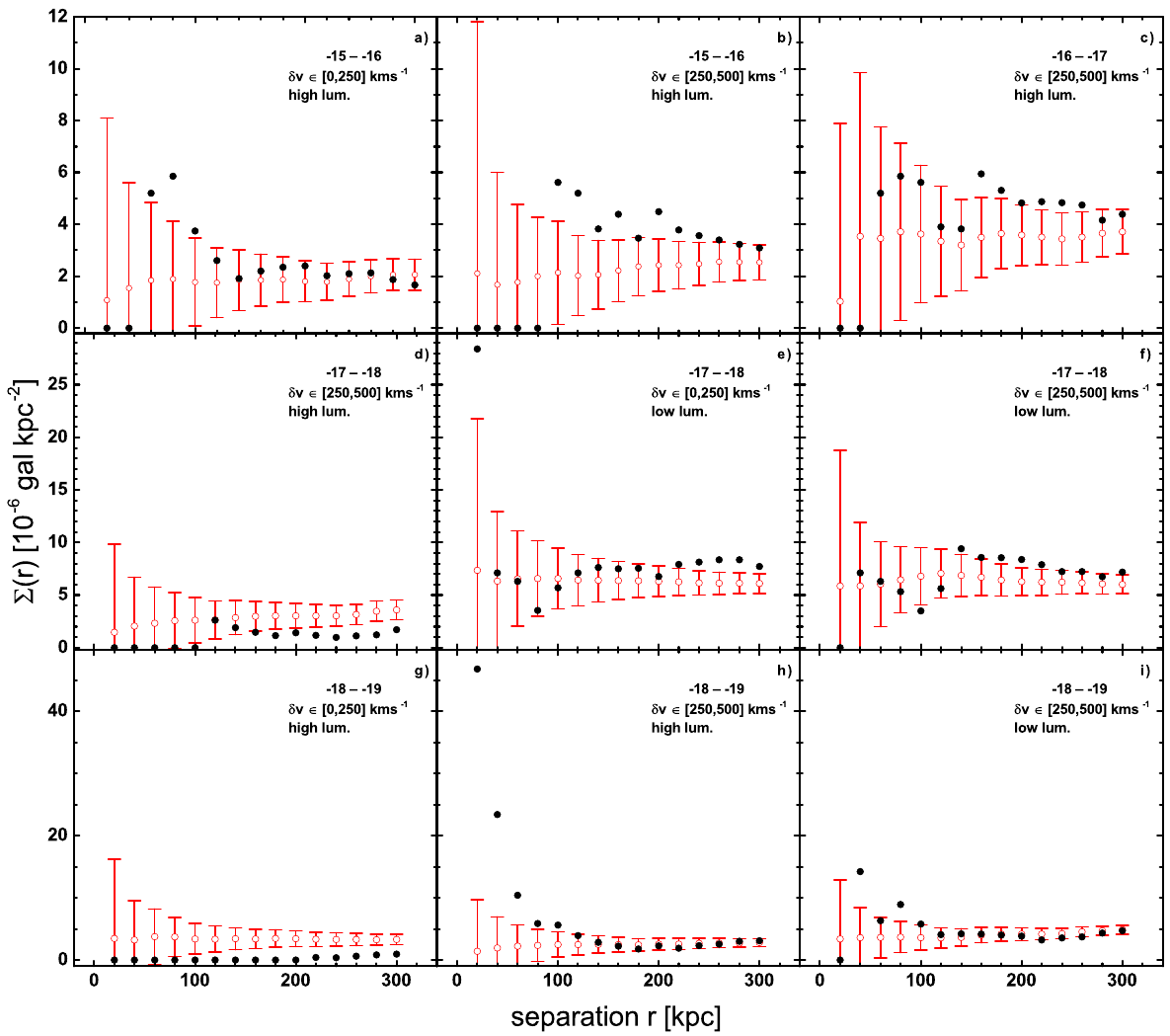}}
\caption{Mean density of different sets of secondaries around high- and low-luminosity primaries depending on separation $r$. Secondaries are distinguished by luminosity (bin of absolute magnitude) and velocity separation, as indicated in the individual panels. Black dots indicate the values for the MA08 set, the open circles and error bars show the mean and standard deviation of 1000 pseudo-clusters.}
\label{fig:Lum_sec}
\end{figure*}

In Fig.~\ref{fig:Lum_E_S} we found hints for bound or missing companions depending on the luminosity of primaries. Now we examine the secondary density around these primary subsets in dependence of secondary luminosity and velocity difference. The results, restricted to interesting parameter combinations, are shown in Fig.~\ref{fig:Lum_sec}. The following features can be noted: 
\begin{itemize}
\item Secondaries with ${M_{B} > -16}$ around high-luminosity primaries show a small excess (over-density) for $r_i \approx 60 - \SI{200}{\kilo pc}$ (Fig.~\ref{fig:Lum_sec}a,b).
\item For secondaries in the next brighter magnitude intervals, ${-17 < M_{B} \leq -16}$ and
${-18 < M_{B} \leq -17}$, there a hint of an excess at large separations, $r_i > \SI{150}{\kilo pc}$ (Fig.~\ref{fig:Lum_sec}c-f). 
\item Secondaries in the brightest magnitude bin, ${-19 < M_{B} \leq -18}$ (Fig.~\ref{fig:Lum_sec}g-i) show the clearest sign for companions at small separations, $r_i <$ 100 kpc, but only for the high velocity difference bin, $\delta v \leq \SI{250}{\km.s^{-1}}$, very significantly around high-luminosty primaries (Fig.~\ref{fig:Lum_sec}h).
\end{itemize}

This last-mentioned Fig.~\ref{fig:Lum_sec}h is essentially reproducing the strong effect seen in Fig.~\ref{fig:Sec_nurE}. Hence the best evidence for bound companions so far is for bright early-type (E/S0) secondaries around bright giant galaxies, in particular luminous spirals, if Fig.~\ref{fig:Lum_E_S} is taken into account.

\subsection{Interaction parameters for the MA08 catalog}
\label{sec:IP}

In Sect.~\ref{sec:IP_Shuffling} we introduced a set of different ‚interaction parameters‘ (IP’s) as an additional tool to look for bound companions. Following F92, $I_{Ferg}$ is used to look for bound pairs of galaxies. $I_{group}$ and $I_{sub}$ were constructed to have additional IP’s that are more sensitive to subgroups around three influential neighbors and to general substructure up to a size of \SI{300}{\kilo pc}, respectively. Finally, $I_{Ka05}$ which is based on tidal interaction is briefly considered as well.

\subsubsection{CDF’s of the interaction parameters}
\label{sec:IP_cdf}

\begin{figure}[h]
\resizebox{\hsize}{!}{\includegraphics[width=0.9\textwidth]{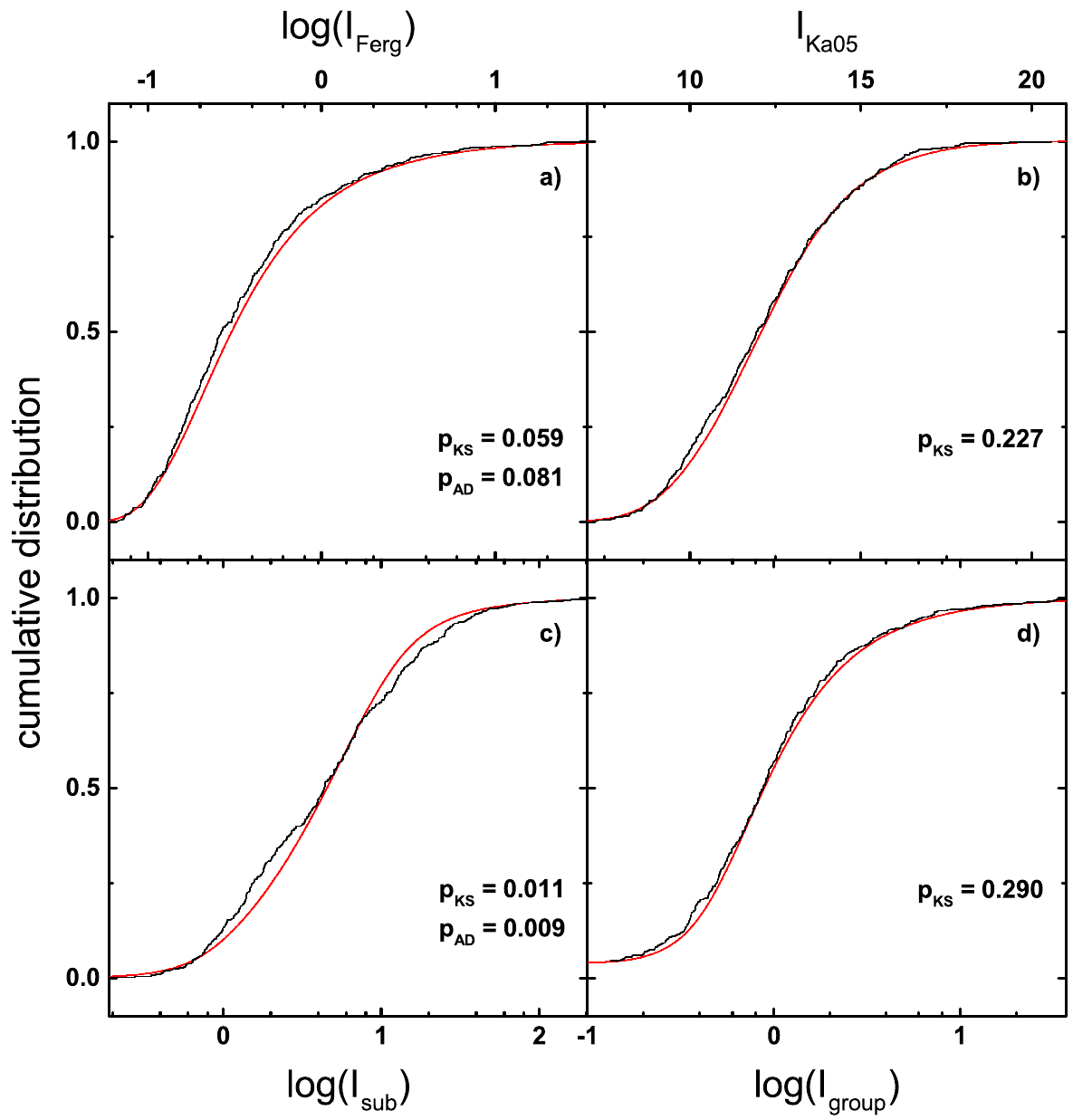}}
\caption[Cumulative distributions of the interaction parameters]{Cumulative distributions for the different interaction parameters discussed in the text. They are indicated along the abscissa. The black curves give the corresponding CDF’s of Coma cluster data from MA08, the red curves show the randomized data for comparison (from galaxy position swapping). The p-values for the statistical tests are indicated in each panel.}
\label{fig:IP}
\end{figure}

It should be remembered that the IP method does not distinguish between primaries and secondaries. The pseudo-clusters are therefore built by direct swapping the positions of randomly chosen galaxies (see Sect.~\ref{sec:direct_pos_Meth}). In Fig.~\ref{fig:IP} we show the cumulative distributions (CDF’s) for all four IP’s. The p-values of the two statistical (KS and AD) tests applied to the CDF’s are also indicated. Obviously, for $I_{Ka05}$ and $I_{group}$ there is no significant difference between real and model cluster (panels b and d), whereas \citeauthor{F92}'s IP is just on the border of statistical significance (panel a). However, $I_{sub}$ is very clearly different for MA08 data and the pseudo-clusters, with a high statistical significance. 
A more detailed discussion is given in the following section using Table~\ref{tab:IP_frac}. First, we perform a test which is only feasible with \citeauthor{F92}'s IP. 

\subsubsection{Bound companions ($I > 2$) and free members ($I < 0.5$)}
\label{sec:IP_2_05}
In the following, galaxies with $I > 2$ are assumed to be gravitationally bound whereas such with $I < 0.5$ are treated as free-floating members of the cluster. Galaxies with $0.5 \leq I \leq 2$ are not statistically evaluated. We focus here on the IP of F92 which is sensitive for pairs. By definition the chosen limits imply that a galaxy is twice or half as strongly bound to another one than to the cluster as a whole. If a very large fraction of galaxies with high IPs are physically bound, then the dispersion of $\delta v$ is expected to be around \SI{250}{\kilo m.s^{-1}} or less according to F92. On the other hand, with no bound companions at all the dispersion should be $\sim \sqrt{2}\sigma_v$ ($\approx \SI{1530}{\kilo m.s^{-1}}$), where $\sigma_v$ is the mean velocity dispersion of Coma ($\sigma_v = \SI{1082}{\kilo m.s^{-1}}$, CD96). In other words, in this case most of the high IP values are caused simply by projection. 

\begin{figure}[htbp]
\centering
{\includegraphics[width=8cm]{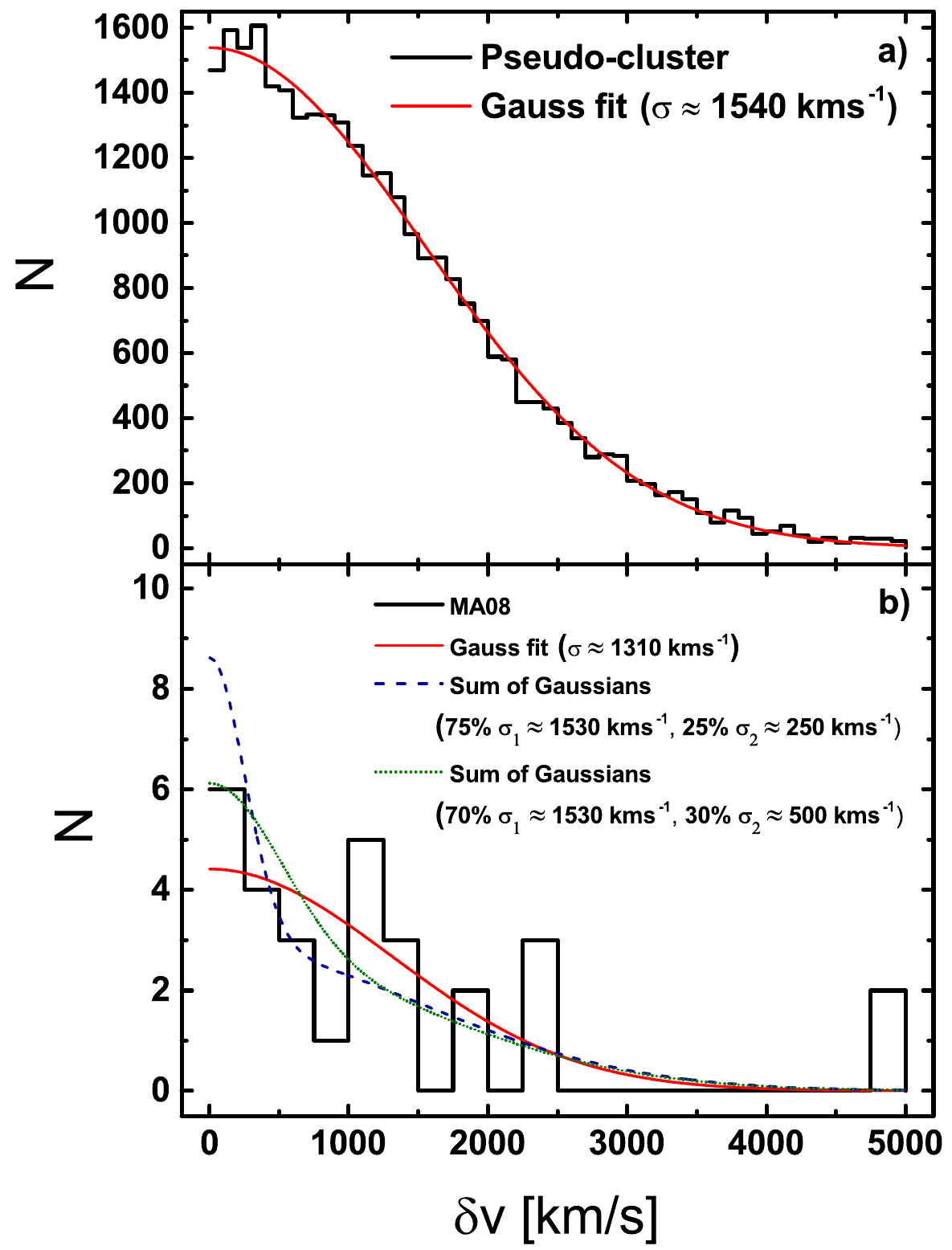}}
\caption[Distribution of $\delta v$ between galaxies with $I_{Ferg} > 2$ and their nearest neighbors]{Histogram of $\delta v$ for the 1000 pseudo-clusters (panel a) and for the $I_{Ferg} > 2$ pairs in the MA08 data (panel b). Both histograms are overlaid with a best-fitting Gauss curve (red). For the MA08 data two combinations of Gauss distributions representing a population of bound pairs and one of projected pairs are drawn as well (dashed and dotted curves in panel b). The area under each curve is always normalized to the total number of galaxies involved.}
\label{fig:delta_v}
\end{figure}

Fig.~\ref{fig:delta_v} shows the distribution of $\delta v$ (up to \SI{5000}{\kilo m.s^{-1}}) between galaxies with $I_{Ferg} > 2$ and their nearest neighbors for the 1000 pseudo-clusters (panel a) and the 29 galaxies with these properties in MA08 (panel b) for which $\delta v$ could be determined. Best-fitting Gauss curves are also shown in the figure. Reassuringly, the pseudo-cluster best-fitting $\sigma_1$ is nearly the same as $\sqrt{2}\sigma_v$ (1540 versus 1530 km $s^{-1}$). 
For the MA08 data the number of galaxies with large values of $I_{Ferg}$ is of course quite small, but to draw a comparison to \citeauthor{F92}'s work (only 45 galaxies, too) his approach is taken here as well. The best-fitting dispersion $\sigma_* = \SI{1310}{\kilo m.s^{-1}}$ of these galaxies can be modeled by a (normalized) superposition of two Gaussians in the following way: Assuming that part of the galaxies are indeed bound in pairs, they should be drawn from a narrow Gaussian with width $\sigma_2 = \SI{250}{\kilo m.s^{-1}}$, while those not bound should follow $\sigma_1 = \SI{1530}{\kilo m.s^{-1}}$ as described above. More formally, the total distribution function $\sigma_*$ as a combination of the two Gaussians is (with $\beta\in[0,1]$):

\begin{equation}
\label{eq:sumGauss}	
\sigma_*^2 = \beta \sigma_1^2 + (1-\beta) \sigma_2^2\,\,\,.
\end{equation}

Solving Eq.~\eqref{eq:sumGauss} for $\beta$ leads to a fraction of roughly 0.75, implying that $\sim 25\%$ of those 29 galaxies with $I_{Ferg} > 2$ are indeed physically bound ($\sim 2\%$ of all 353 studied galaxies). The corresponding best-fitting superposition is shown as dashed blue curve in Fig.~\ref{fig:delta_v} is shown for the MA08 data. This result matches the expectation that only few bound satellites in Coma exist. However, as the galaxy census of Coma is certainly far from being complete, in particular for faint dwarf members which are most prone to be bound companions, this fraction of companions is bound to raise in the future, and $2\%$ must be regarded as a lower limit.

A third (dotted green) curve is shown in the MA08 part of Fig.~\ref{fig:delta_v} modeling a combination where the bound pairs have a high velocity difference of 500 km s$^{-1}$. This is indeed in accord with our previous findings that over-densities are found preferably for velocity differences in the range between 250 and \SI{500}{\kilo m.s^{-1}} (e.g., Fig.~\ref{fig:Sec_nurE}). In this case, the fraction $\beta$ in Eq.~\eqref{eq:sumGauss} becomes $\sim 0.3$ for galaxies with $I_{Ferg} > 2$ drawn from a Gaussian with $\sigma_2 = \SI{500}{\kilo m.s^{-1}}$, which is not very different from the case before. As can be seen in Fig.~\ref{fig:delta_v} all three curves may describe the histogram due to the small number of galaxies. Even a Gaussian curve with $\sim \sqrt{2}\sigma_v \approx \SI{1530}{\kilo m.s^{-1}}$ would fit into the picture. It is therefore clearly impossible to give an exact fraction of bound companions. However, a much larger fraction of bound satellites (e.g., as large as in the Virgo cluster with $7\%$, F92) can almost certainly be excluded.

$I_{sub}$ and $I_{group}$ aim at the gravitational interaction between more than just two neighboring galaxies as $I_{Ferg}$. The above analysis based on velocity differences between a galaxy and its nearest neighbor is therefore not possible for these IPs. The same is true for $I_{Ka05}$, as no limit for bound or free galaxies can be determined easily (s. Sect.~\ref{sec:IP_theory}). Nevertheless, the distribution of the interaction parameters $I_{sub}$ and $I_{group}$ and their dependence on morphology can be studied and compared to the one in 1000 pseudo-clusters in a different way, as follows. 

In Table \ref{tab:IP_frac} we list the fractions (percentages) of bound pairs ($I > 2$) and free-floating cluster members ($I < 0.5$) for $I_{Ferg}$, $I_{sub}$, and $I_{group}$, comparing the real cluster data from MA08 with the random clusters for each morphological type. The numbers referred to in the following discussion are emphasized. The columns give: (1) Morphological type; (2) number and percentage of galaxies with $I > 2$ and $I < 0.5$ for \citeauthor{F92}'s IP in the data and for the randomized clusters; (3) the same for substructure IP and (4) for the group IP. The percentages for the pseudo-clusters are mean values. The standard deviations $\hat{\sigma}$ are similar for all types and amount to $\hat{\sigma} \approx 0.8-0.9\%$ for $I_{Ferg} > 2$, and to $\hat{\sigma} \approx 1.2-1.4\%$ for $I_{sub} > 2$ and $I_{group} > 2$. For the free members the values are $\hat{\sigma} \approx 1.4-1.5\%$ for the IP of F92, $\hat{\sigma} \approx 0.4-0.5\%$ for the substructure IP and $\hat{\sigma} \approx 1.3\%$ for the group IP.

\begin{table*}[htbp]
\small

  \caption{Fractions of bound companions and free members according to the IP-analysis.}
    \begin{tabular}{l>{\raggedleft\arraybackslash}m{2em}>{\raggedleft\arraybackslash}m{2em}>{\raggedright\arraybackslash}m{2em}>{\raggedleft\arraybackslash}m{2em}>{\raggedleft\arraybackslash}m{4em}>{\raggedleft\arraybackslash}m{2em}>{\raggedleft\arraybackslash}m{2em}>{\raggedleft\arraybackslash}m{2em}>{\raggedleft\arraybackslash}m{4em}>{\raggedleft\arraybackslash}m{2em}>{\raggedleft\arraybackslash}m{2em}>{\raggedleft\arraybackslash}m{2em}}
    \hline\hline
       Type   & \multicolumn{4}{c}{$I_{Ferg}$}       & \multicolumn{4}{c}{$I_{sub}$}   &   \multicolumn{4}{c}{$I_{group}$}  \\
          \multicolumn{1}{c}{(1)} & \multicolumn{4}{c}{(2)} & \multicolumn{4}{c}{(3)} & \multicolumn{4}{c}{(4)}\\
    \hline
          & MA08  &       & 1000 pseudo-clusters &       & MA08  &       & 1000 pseudo-clusters &       & MA08  &       & 1000 pseudo-clusters &  \\
   
          & N     & \%    & N     & \%    & N     & \%    & N     & \%    & N     & \%    & N     & \% \\
          \hline
          \multicolumn{13}{c}{$I > 2$} \\
          \hline
    E/S0  & 14    & 9.1   & 13377 & 8.7 & 109   & \textbf{70.8}  & 116291 & \textbf{75.5}  & 32    & 20.8  & 33147 & 21.5 \\
    S     & 8     & 9.1   & 7651  & 8.7 & 54    & \textbf{61.4}  & 66291 & \textbf{75.3}  & 15    & \textbf{17.0}    & 19162 & \textbf{21.8} \\
    dE    & 16    & 8.7   & 15711 & 8.5 & 129   & \textbf{70.1}  & 138516 & \textbf{75.3}  & 34    & 18.5  & 39402 & 21.4 \\
    Irr   & 1     & \textbf{2.1}   & 4092  & \textbf{8.7} & 35    & 74.5  & 35540 & 75.6  & 12    & \textbf{25.5}  & 10132 & \textbf{21.6} \\
\hline
    \multicolumn{13}{c}{$I < 0.5$} \\
    \hline
    E/S0  & 105   & 68.2  & 103762 & 67.4 & 3     & 1.9   & 4311  & 2.8   & 37    & 24.0    & 36793 & 23.9 \\
    S     & 66    & \textbf{75.0}    & 59135 & \textbf{67.2} & 4     & \textbf{4.5}   & 2518  & \textbf{2.9}   & 30    & \textbf{34.1}  & 20939 & \textbf{23.8} \\
    dE    & 127   & 69.0    & 124286 & 67.5 & 3     & 1.6   & 5106  & 2.8   & 45    & 24.5  & 44045 & 23.9 \\
    Irr   & 36    & \textbf{76.6}  & 31630 & \textbf{67.3} & 2     & 4.3   & 1345  & 2.9   & 9     & \textbf{19.1}  & 11239 & \textbf{23.9} \\
    \hline
    \end{tabular}%
  \label{tab:IP_frac}%
\end{table*}%

As can be seen in Table~\ref{tab:IP_frac} (col. 2), the fraction of galaxies with $I_{Ferg} > 2$ in MA08 is marginally higher than in the randomized clusters for all types except the irregulars. In the pseudo-clusters the fraction of each type is approximately the same, for all IPs and for free members as well, which is expected from the direct position shuffling method. The low fraction of bound irregulars may be a small number effect, if true it would contradict our previous findings. The numbers for $I_{Ferg} < 0.5$ are more revealing. Here late-type galaxies, both spirals and irregulars, are significantly overabundant among free-floating members when compared to the pseudo-clusters (at $5\hat{\sigma}$-level).

$I_{sub}$ compares the gravitational pull between a galaxy and its neighbors within \SI{300}{\kilo pc} to the pull it is exposed to by the rest of the cluster. Hence as expected, the percentage of bound (non-bound) galaxies of every type is strikingly larger (smaller) than for $I_{Ferg}$. A systematic deviations between the distributions of $I_{sub}$ in Coma and the pseudo-clusters was already noted in Sect.~\ref{sec:IP_cdf}, Fig.~\ref{fig:IP}c, in the sense that there are more galaxies in the real cluster with small $I_{sub}$ and less with large values compared to the pseudo-clusters. This is confirmed here. Moreover we see that the effect is essentially owed to spiral galaxies. There are considerably fewer late types included in substructure than in the randomized clusters (at $10\hat{\sigma}$-level). Spirals are apparently underabundant in the known substructures within the MA08 area. But even luminous early-types and dwarfs follow this pattern, albeit less strongly (still at $3-4\hat{\sigma}$). Overall, these results taken together are in accord with the view that Coma is a fairly relaxed cluster. 

The group IP is a weaker variant of the substructure IP, describing the three strongest gravitational interactions of galaxies within \SI{300}{\kilo pc}. The values differ notably only for the spirals (again) for $I_{group} > 2$ ($7\hat{\sigma}$) and $I_{group} < 0.5$ ($3\hat{\sigma}$). Another point is that the irregulars experience statistically more interaction with three neighbors than with their nearest neighbor. This seems to indicate that irregulars are rather bound to subgroups (infalling groups) rather than single primaries. 


\subsection{Results for the $\Sigma(r)$-analyses of the WLG11 catalog}
\label{sec:WL11_results}

The settings for the MC simulations analyzed in this section are almost the same as in Sect.~\ref{sec:MA08_results}, except for the following changes: 1) the primary-secondary discrimination is at $M_{R}=-20$ (228 giants and 695 possible companions), and 2) absolute magnitudes can be taken directly from the catalog of WL11.

Again primaries with minor axis (referring to the position shuffling, s. Eq.~\eqref{a-b-ratio2}) smaller than \SI{250}{\kilo pc} are excluded. A second cut is applied to avoid density correction due to boundary effects (s. Sect.~\ref{sec:WL11_sample}), leaving 197 primaries in the sample. A primary of this set is not necessarily a primary in the MA08 analysis and vice versa. This time all secondaries have velocity data (thanks to SDSS). The velocity bins of primary-secondary $\delta v$ will be the same as before.
Position shuffling is found to produce constantly lower random-densities than the $v$-shuffling method. To be conservative, we therefore rely on, and show here only, the results obtained with the velocity randomization method. The prize, however, is that no statistical tests can be executed, as the galaxy separations remain the same with this method. 

\subsubsection{Complete set with $\delta v$ distinction}
\label{sec:WL11_All}
\begin{figure}[h]
\centering
\includegraphics[width=1\linewidth]{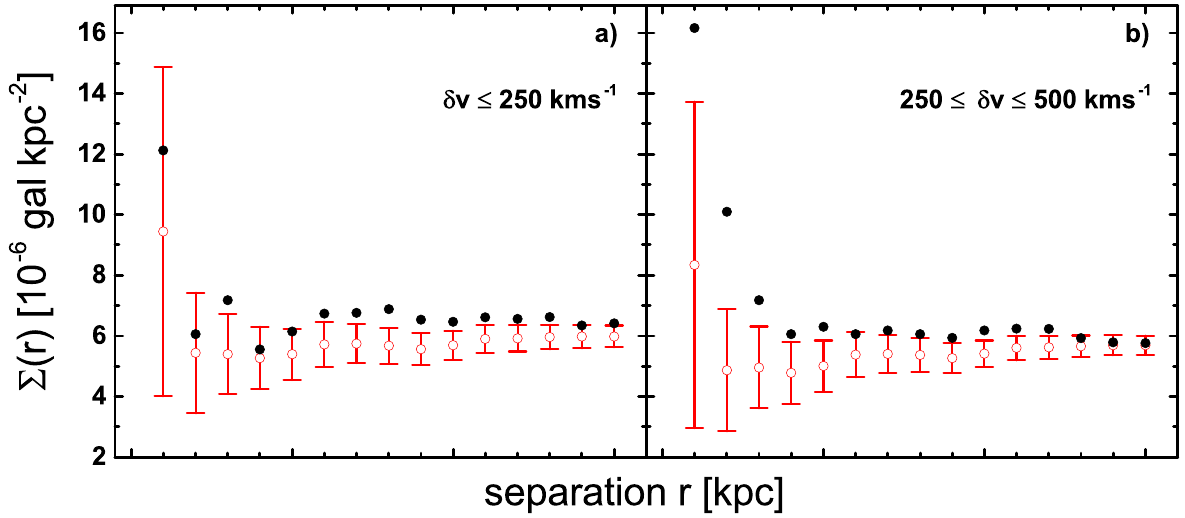}
\caption[Mean density of secondaries around primaries for the WLG11 data]{Mean density of secondaries around primaries for the complete WL11 set, separated into two bins of velocity difference as indicated in the panels. Black dots represent the data; open circles and error bars in red give the mean and standard deviation of 1000 pseudo-clusters created by shuffling primary velocities.}
\label{fig:WL11}
\end{figure}

Fig.~\ref{fig:WL11} provides a first look at the complete WLG11 sample. The only differentiation made is by primary-secondary $\delta v$, divided into the usual two bins. There is a clear excess of possible bound companions apparent, in particular (again), at the $2\hat{\sigma}$-level, for high velocity differences ($250 < \delta v \leq \SI{500}{\km.s^{-1}}$).

In the following we elaborate this result with respect to primary position, and primary and secondary luminosity. A distinction by type cannot be done, as WGL11 lacks morphological information. 

\subsubsection{Dependence on location in cluster and luminosity}
\label{sec:WL11_Diff}
As a first step the primary sample is divided into a number of elliptical annuli around the cluster center.
For a better comparison with the results for the MA08 catalog, the first separation is again fixed at minor axis length $b$=\SI{250}{\kilo pc} and named inner and outer as in Sect.~\ref{sec:250kpc_PrimMorph}. The outer region is further subdivided into two distance regions of width \SI{0.25}{\mega pc} out to 1\,Mpc, and 5 regions of width \SI{0.5}{\mega pc} out to 3.5\,Mpc. The sets are given in Table~\ref{tab:WL11_loc}. The columns list the following: (1) name of the set, (2) range for the primary minor axis $b$, (3) number of galaxies in each subsample. The whole set of elliptical annuli is also shown in the Coma cluster map of WGL11 data, Fig.~\ref{fig:Karte_WL11}.

\begin{table*}
  \centering
  \caption[Local grouping of WLG11 primaries (brighter than $M_{R}=-20$)]{Distance grouping of primary galaxies (brighter than $M_{R}=-20$)}
    \begin{tabular}{lp{16em}l}
    \hline\hline
    Set ID & Minor axis range [\SI{}{\mega pc}] & Number of galaxies \\
    (1) & (2) & (3) \\
    \hline
    
    inner  & ]0.25,0.5] & 31 \\
    
    outer     & ]0.5,0.75] & 24 \\
    
    outer II & ]0.75,1] & 17 \\
    
    outer III & ]1,1.5] & 31 \\
    
    outer IV & ]1.5,2] & 34 \\
    
    outer V & ]2,2.5] & 20 \\
    
    outer VI & ]2.5,3] & 26 \\
    
    outer VII & ]3,3.5] with restriction of $a < \SI{3.9}{\mega pc}$ & 14\\ 

    \hline
    \end{tabular}%
  \label{tab:WL11_loc}%
\end{table*}

The results are shown here in form of a table instead of the usual density distribution: see Table~\ref{tab:WL_Loc}. The columns list: (1) primary subsample according to Table~\ref{tab:WL11_loc}, (2) velocity difference bin, (3) net secondary density around primaries (WL11 data minus random values). For each bin $[0,r_i]$ we give either a "0" or a series of "+" or "-" signs in the table. A "0" indicates that the WL11 mean density $\Sigma(r)$ lies within one standard deviation of the pseudo-cluster mean $\hat{\Sigma}(r)$. One, two or three "+"-signs are indicating a $1\hat{\sigma}$, $2\hat{\sigma}$ or $3\hat{\sigma}$ over-density. The latter also stands for over-densities at more than the $3\hat{\sigma}$ significance level. Similarly, under-densities are indicated by "-"-signs.

\begin{table*}[htbp]
\tiny
    \caption[Net densities of secondaries around primaries depending on their location for the WLG11 data]{Net densities $\hat{\Sigma}_{net}(r)$ of secondaries around primaries depending on their location according to Table~\ref{tab:WL11_loc}. Symbols are explained in the text.}
    \begin{tabular}{p{5em}p{3.5em}rrrrrrrrrrrrrrr}
    \hline\hline
    Primary \newline Location & $\delta v$-bin \newline [\SI{}{\kilo m.s^{-1}}] & \multicolumn{15}{c}{$\hat{\Sigma}_{net}(r)$} \\
    (1) & (2)   & \multicolumn{15}{c}{(3)} \\ \hline
\multicolumn{2}{c}{Separation r [\SI{}{\kilo pc}]}                & 20 & 40 & 60 & 80 & 100 & 120 & 140 & 160 & 180 & 200 & 220 & 240 & 260 & 280 & 300 \\
           \hline
    Outer II & 0-250 & 0     & 0   & ++     & 0     & +     & +     & 0     & +     & 0     & 0     & 0     & 0     & 0     & 0     & 0 \\
    & 250-500 & 0     & ++     & +     & +     & +     & 0     & 0     & 0     & 0     & 0     & +   & +   & 0     & 0     & 0 \\
     \hline
    Outer III & 0-250 & 0     &++   & +     & 0     & 0     & +     & 0     & 0     & 0     & +     & +     & +     & +     & +     & 0 \\
    & 250-500 & 0     & 0     & +     & 0     & +     & ++     & +     & +     & +     & +     & +   & ++   & ++     & ++     & ++ \\
     \hline
    Outer IV & 0-250 & 0     & +     & 0   & 0   & 0   & 0     & +     & 0     & +     & +   & +   & +   & +     & +     & + \\
          & 250-500 & 0     & 0     & 0     & +     & +     & ++     & ++     & ++     & +     & ++   & +   & +   & +   & +   & + \\
           \hline
    Outer V & 250-500 & 0     & + +   & 0     & 0     & 0     & 0   & 0     & 0     & 0     & 0     & 0     & 0     & 0     & 0     & 0 \\
     \hline
    Outer VI & 0-250 & +     & +     & 0     & 0     & 0     & 0     & 0     & +     & 0     & 0     & 0     & +     & +     & +     & + \\
          & 250-500 & 0     & 0     & 0     & -     & -     & 0     & -     & -     & -     & 0     & 0     & 0     & -     & -     & 0 \\
    \hline
    Outer VII & 0-250 & 0     & 0     & 0     & 0     & +     & ++     & +     & +     & +     & +     & +     & +     & +     & 0     & 0 \\
          & 250-500 & +++     & +++     & +     & +     & +     & 0     & +     & +     & +     & +     & 0     & 0     & 0     & 0     & 0 \\
    \hline
    \end{tabular}%
  \label{tab:WL_Loc}%
\end{table*}%

Not given are the (null) results for the inner and next outer regions where no clear over-densities for the complete sample were found for the WL11 sample either. But Table~\ref{tab:WL_Loc} confirms that over-densities (bound companions) abound in the outskirts of the cluster, and again more so in the high-$\delta v$ bin. To cover the outskirts of the cluster was of course the motivation to work with WGL11 data in addition. We cannot aim at interpreting the results for the individual distance regions, however. They are lumped together again in the following.  

The primary sample is split into a 63 high- ($M_R < -21$) and 130 low- ($-21 \geq M_R \geq -20$) luminosity galaxies. For the high-luminosity primaries we find $1\hat{\sigma}$-over-densities of secondaries in the innermost \SI{60}{\kilo pc} separation range. For low-luminosity primaries
there are $1-2\hat{\sigma}$-level over-densities for separations larger than $\sim \SI{100}{\kilo pc}$ for $\delta v \leq \SI{250}{\kilo m.s^{-1}}$, and over-densities at the same significance level over almost the whole range of $r_i$ for the higher $\delta v$-bin. We show these results in combination with a distinction by secondary luminosity in the following.

\begin{figure}[htbp]
\centering
\includegraphics[width=1\linewidth]{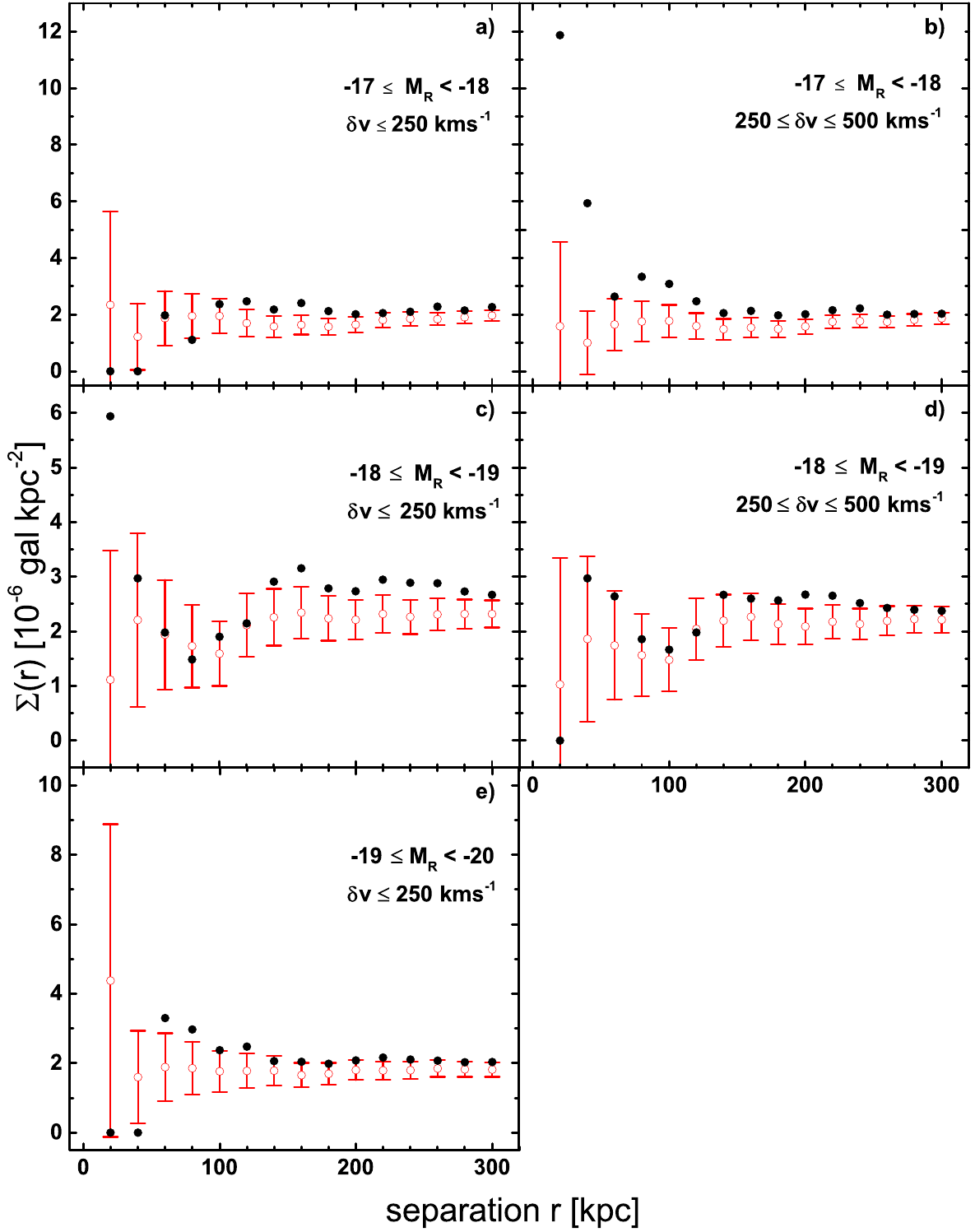}
\caption[Mean density of different secondary subsets around low luminous primaries]{Mean density of secondaries around low-luminosity primaries. Secondary luminosity and velocity difference bin are indicated in each panel. Black dots: WL11 data; open circles and error bars in red: mean and standard deviation of 1000 pseudo-clusters created by primary velocity shuffling.}
\label{fig:WL11_Lum_Sec_Low}
\end{figure}

\begin{figure}[htbp]
\centering
\includegraphics[width=1\linewidth]{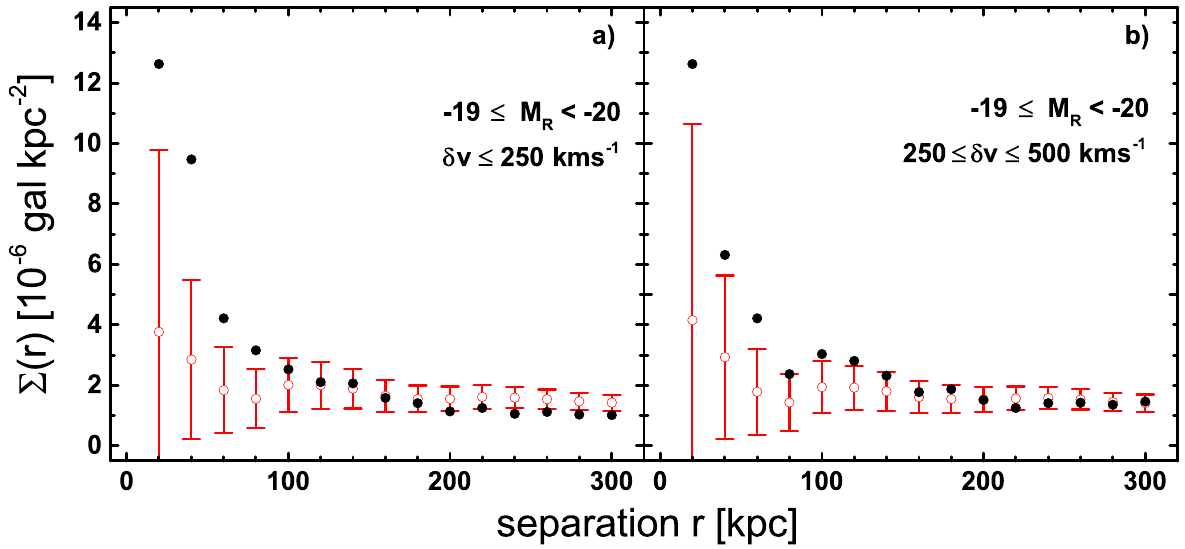}
\caption[Mean density of secondaries with $-20 < M_{R} \leq -19$ around high luminous primaries]{Mean density of the brightest secondaries around high-luminosity primaries. Otherwise like Fig.~\ref{fig:WL11_Lum_Sec_Low}.}
\label{fig:WL11_Lum_Sec_High}
\end{figure}

The secondaries are divided into three luminosity groups: 213 galaxies with $-20 < M_{R} \leq -19$, 257 with $-19 < M_{R} \leq -18$, and 225 with $-18 < M_{R} \leq -17$. The resulting secondary densities, restricted to cases where a noteworthy signal is present, are shown in Fig.~\ref{fig:WL11_Lum_Sec_Low} for secondaries around low-luminosity primaries, and in Fig.~\ref{fig:WL11_Lum_Sec_High} for secondaries around high-luminosity primaries.

We note that secondaries of all luminosities show a mild over-density over almost all separations, where the luminosity class $-18 < M_{R} \leq -17$ pops up with very significant over-densities at separations below ca. 100 kpc for high velocity differences (Fig.~\ref{fig:WL11_Lum_Sec_Low}b). The over-densities, again at small separations, around high-luminosity primaries mentioned above are clearly owed to the most luminous secondaries, in the low- and high-$\delta v$ bin alike (Fig.~\ref{fig:WL11_Lum_Sec_High}).  

\subsubsection{Interaction parameters for the WLG11 catalog}
\label{sec:WL11_IP}

Repeating the IP analysis from Sect.~\ref{sec:IP} for the WL11 sample, we first find a notable difference in that the CDF for $I_{sub}$ (compare Fig.~\ref{fig:IP}) is no longer significantly different between data and random model ($p_{KS} = 0.262$). 

\begin{figure}[htbp]
\centering
{\includegraphics[width=8cm]{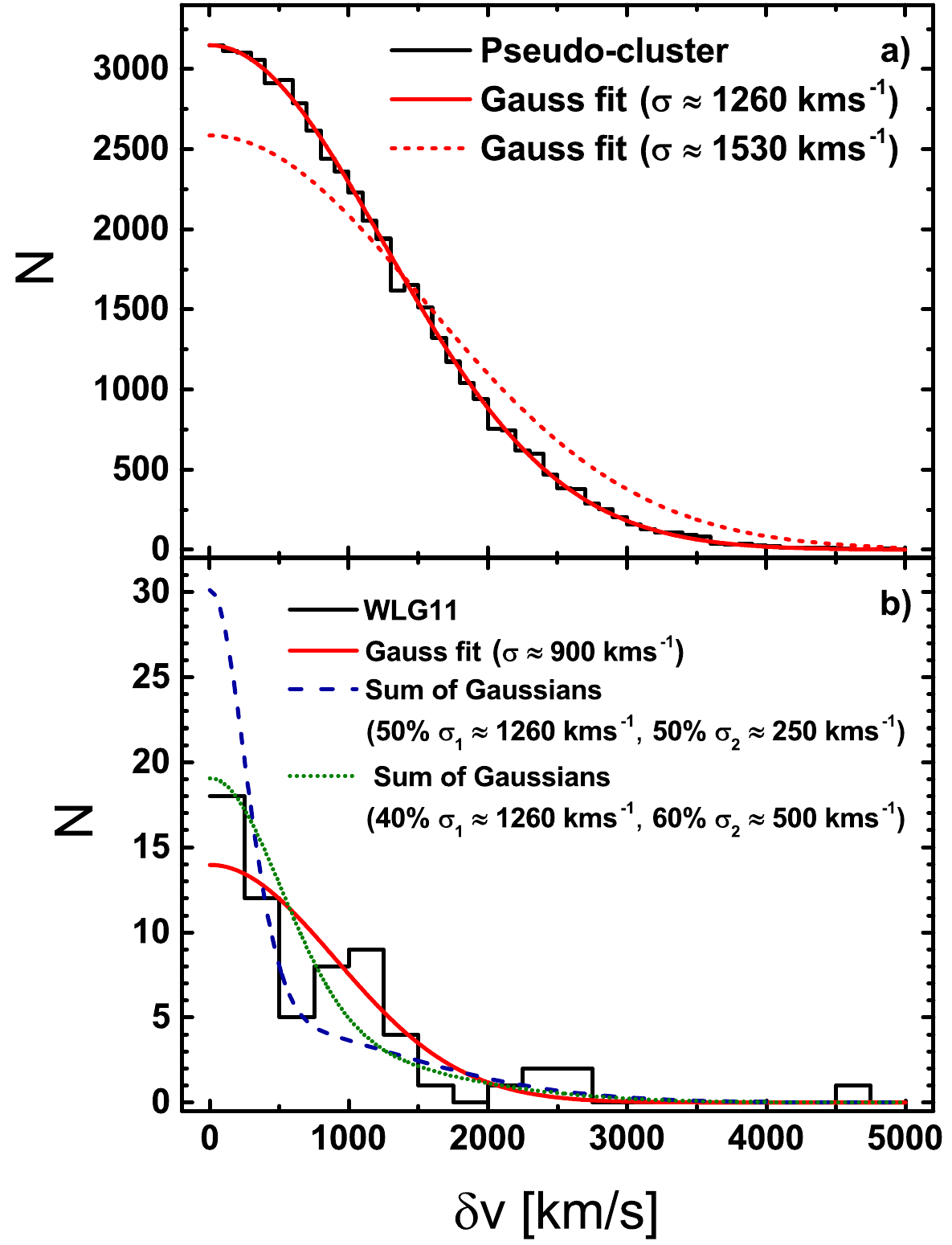}}
\caption[Distribution of $\delta v$ between galaxies with $I_{Ferg} > 2$ and their nearest neighbors]{Histogram of $\delta v$ for the 1000 pseudo-clusters (panel a) and for $I_{Ferg} > 2$ pairs in the WL11 data (panel b). Both histograms are overlaid with a best-fitting Gauss curve (red). For the pseudo-clusters a Gaussian with the velocity dispersion given by CD96 is shown for comparison (dotted line in panel a). The observed WL11 histogram is modeled by two different combinations of Gauss curves as explained in the text (dashed and dotted curves in panel b). The area under each curve is always normalized to the number of galaxies.}
\label{fig:WL11_delta_v}
\end{figure}

In addition, the best Gaussian fit to the $\delta v$-distribution for pseudo-clusters leads to a smaller value for the cluster dispersion $\sigma_v = \frac{\SI{1260}{\kilo m.s^{-1}}}{\sqrt{2}} \approx \SI{890}{\kilo m.s^{-1}}$ than before (see Fig.~\ref{fig:WL11_delta_v}a). For comparison, the dotted line in Fig.~\ref{fig:WL11_delta_v}a shows a Gaussian based on $\sigma_v = \SI{1080}{\kilo m.s^{-1}}$ as used in Sect.~\ref{sec:IP_2_05} \citep{CD96}. All commonly used dispersion values for Coma are in the range $\sigma_v \approx 1000 - \SI{1100}{\kilo m.s^{-1}}$. \cite{Edw02} found a lower value of $\sigma_v \approx \SI{980}{\kilo m.s^{-1}}$ but only for the giant population. The velocity dispersion of the dwarf population is even higher than the overall mean (e.g., \SI{1096}{\kilo m.s^{-1}} by \cite{Edw02} or \SI{1213}{\kilo m.s^{-1}} by \cite{Chi10}). However, the explanation for this apparent discrepancy probably lies in the observation that the low-density outskirts of a cluster are dynamically cooler and hence exhibit smaller velocity dispersion than the central part. 
Studying infalling substructures, A05 found a velocity dispersion of $\sigma_v \approx \SI{973}{\kilo m.s^{-1}}$ for the whole cluster, still being on the high side but going in the right direction.

The IP analysis can be performed in any case. Fitting a Gaussian to the distribution of $\delta v$ for the 63 $I_{Ferg} > 2$ pairs in the WL11 data leads to $\sigma_* \approx \SI{900}{\kilo m.s^{-1}}$ (see Fig.~\ref{fig:WL11_delta_v}b). Now we can again decompose this distribution into a component of projected pairs with $\sigma_1 \approx \SI{1260}{\kilo m.s^{-1}}$ (obtained from the pseudo-clusters) and a component of bound companions with either $\sigma_2 = \SI{250}{\kilo m.s^{-1}}$ or $\sigma_2 = \SI{500}{\kilo m.s^{-1}}$. Using Eq.~\eqref{eq:sumGauss} one finds $\beta \approx 50\%$ for the former case and $\beta \approx 40\%$ for the latter. These two solutions for combined Gaussians are overlaid in Fig.~\ref{fig:WL11_delta_v}b. The combination with the narrow $\delta v$ component clearly does not provide a good fit and one should favor the second solution with $\sigma_2 = \SI{500}{\kilo m.s^{-1}}$. In this case Fig.~\ref{fig:WL11_delta_v}b suggests that $\sim 60\%$ of pairs with $I_{Ferg} > 2$ are bound companions. Thus around $4\%$ of the 883 galaxies in the WL11 set may be considered bound companions, not necessarily as satellites of an individual primary but more likely of a subgroup.

\section{Discussion}
\label{sec:discussion}

Building on the work of \cite{F92}, several statistical methods have been used to search for bound companions in the Coma cluster. The results are briefly summarized and discussed in this section. Some aspects were discussed already before along with the presentation of the results, especially for the IP analyses.

\subsection{Comparison of mean secondary density around primaries}

\subsubsection{MA08 catalog}
\label{sec:dis_MA08}
We first determined the mean density of secondary galaxies ($M_B > -19$) around primary galaxies ($M_B < -19$) for the MA08 catalog and compared it to the mean of 1000 pseudo-clusters created by shuffling the velocities or positions of the primaries. We analyzed the sample with and without using redshift data (bins for the primary-secondary velocity difference $\delta v$ are: $[0,250]$ and $]250,500]\SI{}{\kilo m.s^{-1}}$). 

A general outcome of this statistical treatment was the frequent or even dominant occurrence of under-densities (Figs.~\ref{fig:8Var}-\ref{fig:Lum_sec}), which are difficult to interpret in terms of physical effects. Boundary effects cannot be blamed for this, as the the Monte Carlo clusters are run within the same boundaries as the real cluster. More likely, these under-densities are an artefact from the simple random shuffling which assumes that the galaxies are distributed randomly in a smooth, single-component cluster potential, on top of which we would have those companions around single primary galaxies. But we know of course from substructure studies of Coma \citep[e.g.,][]{Biviano96} that the galaxies are clustered on all scales in a self-similar way, as is expected from hierarchical clustering. The resulting under-densities are then also expected to be particularly strong in the dense, inner part of the cluster covered by the MA08 catalog, even though the very innermost core region was avoided here for possible confusion. Indeed, in the outer, low-density parts covered by the WLG11 no such under-densities appear (cf. Figs.~\ref{fig:WL11}-\ref{fig:WL11_Lum_Sec_High}). In any case, under-densities, if they are not highly significant, will in general not be taken at face value here. 

Although the ‚companion signals‘ are generally only of low significance (at $1\hat{\sigma}$ for at least three bins in a row) and can be statistically backed with a KS or AD test only in a few cases, some of the findings are in accord with previous observations and are strong enough to be at least indicative of various evolutionary effects that are expected to play a role in a cluster environment. The most noteworthy observations are the following. 

\begin{enumerate}

\item Over-densities around high-luminosity late-type (spiral) galaxies are detected for small and medium separations (up to $\sim \SI{160}{\kilo pc}$) at a significance level of $1\hat{\sigma} - 3\hat{\sigma}$ for five bins (cf.\,Fig.~\ref{fig:Lum_E_S}c). 
This is where we expected to find satellite galaxies. \cite{And96} describes a morphology-velocity segregation, with the late-type galaxy population of the cluster having a higher velocity dispersion ($> 700-\SI{750}{\kilo m.s^{-1}}$) than the early-type one ($\sim 700-\SI{750}{\kilo m.s^{-1}}$). According to CD96 this likely means that late-type cluster galaxies are still in a stage of infall into the cluster core. Our finding that some bound companions still exist around the spirals clearly supports this scenario.  

\item Slight over-densities are found around low-luminosity early-type galaxies at medium separations and high velocity differences (see Fig.~\ref{fig:Lum_E_S}b). In the scenario just mentioned, where early types would be the oldest and therefore most virtualized cluster members, dwarfs bound to individual early-type giants are not expected to exist, with the exception of extremely close companions (as discussed below). More conceivably they are bound to small subgroups. Several groups in the Coma region with a velocity dispersion of $\sim \SI{300}{\kilo m.s^{-1}}$  were identified by A05, but only a few of them are located within the MA08 area. As these authors give no detailed description of the group members, we cannot judge whether the excesses we found is due to those groups. 

We note that points 1 and 2 generally agree with the old finding by \cite{BothunSull} that companions are more numerous around spirals than E/S0’s, although that sample encompassed preferentially isolated galaxies.

At this point we have to mention the Coma cluster study of \citet{SHP97} who found an excess of low-surface brightness dwarfs (dEs) around high-luminosity early-type giants, claiming that an E/S0 giant has on average 4$\pm$1 dwarf companions. This study cannot be directly compared with our’s, however, for a number of reasons. First, \citeauthor{SHP97} confined their dwarf search to the dense core region which we cut out to avoid confusion. Second, their (numerous!) low SB dwarfs are much fainter than our secondaries. Third, bound companions were searched for only in a very small circle ($r <$ 20\,kpc) around the giants (this was their way to avoid confusion). The method employed the modeling of the light distribution of a giant and then subtracting the model from the image. The number of companions was statistically assessed by number background subtraction. In our study we are essentially blind for such extremely close companions that are plausibly bound to an individual giant even in the very dense core region. However, the following two observations are well in accord the \citet{SHP97} finding.  
 
\item An excess of E/S0 secondaries is revealed for higher velocity differences (see Fig.~\ref{fig:Sec_nurE}). This is expected from the morphology-density relation of galaxies in general \citep[e.g.,][]{D80A,D80B} and for the Coma cluster in particular \citep{And96}. Strangely, the over-density is not seen for small $\delta v$ (Fig.~\ref{fig:Sec_Morph})

\item An excess of faint ($M_B > -16$) companions is found around high-luminosity giants for velocity differences between primary and secondary that are indicative of gravitational binding for massive giants (cf.\,Fig.~\ref{fig:Lum_sec}a,b). Again, this is expected, as faint dwarfs dominate among satellites simply due to the luminosity function of galaxies. The excess is in accord with the kinematically blind over-densities found around luminous spirals under point 1, and with the very close companions around E/S0 giants found by \citet{SHP97}, as mentioned above. Unfortunately, a morphological splitting of our small kinematic sample was not feasible.  

\item Finally, for brighter secondaries we observe strong under-densities around high-luminosity primaries at small $\delta v$ (Fig.~\ref{fig:Lum_sec}d,g), but not so at larger $\delta v$ where we find significant over-densities (Fig.~\ref{fig:Lum_sec} h). Notwithstanding the general problem with under-densities (see above), it is tempting here to interpret this effect in terms of galactic cannibalism: According to Chandrasekhar’s formula for dynamical friction \citep[$F_{dyn.fric.} \sim \mathcal{M}^2 v_M^{-2}$,][p. 644]{BinneyTremaine}, more massive, slowly moving galaxies tend to be more affected by cannibalism than less massive, faster moving ones. However, this explanation might be too simple and other interactions among the galaxies and the cluster more important.

\end{enumerate}

\subsubsection{WLG11 catalog}
\label{sec:WL11disc}
As the Coma cluster clearly exceeds the area covered by MA08, the catalog of \citep[][without morphological information, however]{WL11} was analyzed as well. 
Interestingly, no faint companions at small velocity differences are detected around 
high-luminosity giants in this catalog. There is a slight over-density of secondaries at small separations for high velocity differences (Fig.~\ref{fig:WL11}), probably due to the very significant over-density of modestly faint secondaries around low-luminosity primaries (Fig.~\ref{fig:WL11_Lum_Sec_Low}b). On the other hand, there is also an over-density of luminous secondaries around the most luminous primaries, for small and high $\delta v$ (Fig.~\ref{fig:WL11_Lum_Sec_High}). Without further morphological information it is difficult to make sense out of this. But given the large mean distance of the WL11 galaxies from the cluster core, the presence of companions around low-luminosity, less massive primaries might indicate the expected absence of strong tidal forces out there.

Splitting the outer cluster region into a number of elliptical annuli (Fig.~\ref{fig:Karte_WL11} and Table~\ref{tab:WL11_loc}), we find significant excesses of secondaries at different separations $r$ and mostly in both $\delta v$-bins for regions II-IV and the outermost region VII (Table~\ref{tab:WL_Loc}). The excess in the outer II region (especially for large values of $\delta v$) might be caused by the presence of the groups around spirals NGC 4921 and NGC 4911 \citep{N03}, again supporting the scenario of infalling groups centered on late-type giants mentioned under point 1 above.
The large substructure around the cD galaxy NGC 4839, on the other hand, might explain the over-densities at high velocity differences in the outer III/IV regions. Even though this primary lies within the outer III area, the group exceeds the boundary to the next outer area. It can be seen for instance in \cite{N01} that the X-Ray area is stretched in the SW direction. Some primaries might fall from the outer IV region into this group having still some bound companions. Moreover, A05 have located some more groups in this region as can be seen in their \mbox{Fig.~\ref{fig:Lum_E_S}}.  

\subsection{Interaction parameter method}

Aside from \citeauthor{F92}'s (\citeyear{F92}) interaction parameter, $I_{Ferg}$, which is a measure of the binding of a galaxy to its nearest neighbor in relation to the gravitational influence from the rest of the cluster, we introduced also a substructure and a group IP, $I_{sub}$ and $I_{group}$, which are more sensitive to the binding of a galaxy to a larger mass aggregate (of up to a scale of 300 kpc in the case of $I_{sub}$). The cumulative distributions for these IP’s are essentially the same for the real MA08 sample and a randomized pseudo-cluster, with the exception of $I_{sub}$ where a marginally significant difference is found (Fig.~\ref{fig:IP}). This meets the expectation that the Coma cluster is primarily a smooth, relaxed cluster with only moderate (though still evident) substructure. It should be remembered that the infalling groups around NGC 4839 \citep{N01} and NGC 4921 and 4911 \citep{N03} are not or only partly covered by the MA08 field. Defining galaxies with $I > 2$ as bound (to a neighbor galaxy or a subgroup) and $I < 0.5$ as free (bound only to the whole cluster), and differentiating according to morphological type, we find that more late-type (spiral) galaxies than the average type are free ($I < 0.5$), or fewer are bound, when compared to a mean randomized pseudo-cluster (cf. Table~\ref{tab:IP_frac}). However, this is probably simply mirroring the general morphology-density relation and, in lack of a luminosity differentiation, is not in conflict with our previous finding that late-type giants (luminous spirals), surmised as being the centers of infalling groups, tend to have companions.   
 
The most important use of the interaction parameter method introduced by F92 is the assessment of the statistical abundance of bound companions in a cluster. By analysing the distribution of velocity differences between galaxies with $I_{Ferg} > 2$ and their nearest neighbors we could estimate that roughly $25\%$ or $30\%$ of these (29 cases) are genuinely bound companions, assuming a typical $\delta v$ of 250 or 500 km\,s$^{-1}$, respectively; the remaining pairs would be projection cases. This amounts to $\approx$ 2$\pm 1\%$ of the total cluster population. Given the brightness limitation of the MA08 catalog ($M_B \lessapprox -15$), this percentage may be underestimated, as their could be more bound satellites among the fainter dwarf cluster members not enclosed in the catalog. Additionally, as explained above, \citet{SHP97} reported evidence for a mean number of very close dwarf satellites around E/S0 giants of 4$\pm$1 per giant. We note, however, that this would add at best another $\lessapprox 2\%$ bound companions (10 giants with 4 satellites among 2250 dwarfs).

Going through the same procedure with the WL11 sample which covers a larger area (at a somewhat fainter magnitude limit), the estimated number of bound companions among cluster members is $\approx 4 \pm 1\%$. Again this might be slightly underestimated for the reasons given above. However, overall we regard a relative abundance of $2-4\%$ bound companions in the Coma cluster, down to a limiting magnitude of $M_B \approx -15$ and disregarding extremely close companions hidden in the high-surface brightnesss central part of the primaries, as a realistic, robust estimate. This must be directly compared to \citeauthor{F92}'s (\citeyear{F92}) estimate of $7\%$ for the Virgo cluster. It should be noted that the limit of absolute magnitude of the redshift sample used for the IP analysis is roughly the same for the Virgo cluster in spite of its proximity. Hence the abundance of bound companions in Coma is about half of that in Virgo. This is in accord with the expectation we have for a regular, more relaxed cluster such as Coma versus a irregular, less relaxed cluster such as Virgo.      
We regard this as the principal outcome of our study.

\section{Conclusion}
\label{sec:concl}

\citet{F92} created pseudo-clusters and compared them to the Virgo cluster in order to search for bound companions around giants. In this study we have applied his methods to the Coma cluster. The catalog of \cite{MA08} was used for a detailed analysis of a $\sim 0.73 \times \SI{0.84}{\deg}^2$ field centered around NGC 4889 and 4874. As expected, we find fewer companions in Coma than in Virgo due to the different evolutionary states of the two clusters, Coma being more regular and relaxed than Virgo. Introducing an interaction parameter, we estimate 2 - 4$\%$ of all Coma members to be bound satellites, as compared to $7\%$ found in Virgo found by \citet{F92} for an equivalent luminosity and separation range. 

The mean surface density $\Sigma(r)$ of galaxies fainter than $M_B = -19$ around primaries (brighter than $M_B = -19$), corrected for the expected mean from pseudo-clusters, was analyzed for various parameter constraints with respect to luminosity and morphology of primaries and secondaries, as well as velocity difference. Density excesses, interpreted as possible companions, are mainly found around very luminous late-type galaxies (spirals), preferentially at large cluster-centric distances. This is in accord with the infall scenario of cluster formation. Moreover, we find that only the faintest dwarfs ($-16 < M_B$) seem to be satellites of individual primaries. Brighter secondaries might have been accreted to the substructures around the central dominant galaxies, or simply added to the general cluster potential. There are also hints of galactic cannibalism in the cluster. 

The same statistical tools are used to search for bound companions in the sample of \cite{WL11} which covers a larger region (circled area out to $\sim \SI{4.2}{\mega pc}$) but has no morphological information. Clear excesses of secondaries (fainter than $M_R = -20$) are visible for almost all regions except for the innermost $\sim \SI{750}{\kilo pc}$. 

Summarizing, also in Coma some bound companions exist but the fraction is clearly lower than in Virgo. The methods used are suitable for a search for companions in Coma and could be applied also to other galaxy clusters. An important requirement for such a task is the existence of a cluster catalog covering a large area and a sufficiently large range of galaxy magnitudes. Also morphological assignments and measured radial velocities for most of the member galaxies should be available. 

The results obtained are often difficult to interpret and do not provide much new physical insight without a more detailed cluster modeling. It might be rewarding to apply \citeauthor{F92}'s method also to morphologically and kinematically detailed $N$-body clusters in various evolutionary stages drawn from cosmological simulations and compare the outcomes with the results found by \citep{F92} for Virgo and by the present study for Coma. 

\begin{acknowledgements}
We thank Thorsten Lisker for making the WLG11 data available to us. We thank an anonymous referee for helpful comments. BB is grateful to the Swiss National Science Foundation for financial support.     
\end{acknowledgements}

\bibliographystyle{aa}
\bibliography{arxiv.bbl}

\end{document}